\documentclass[prb, preprint]{revtex4}
\usepackage{graphicx}
\usepackage{amsmath}
\usepackage{amssymb}
\usepackage{color}
\usepackage{braket} 
\usepackage{etoolbox} 
\usepackage{letltxmacro} 
\usepackage{xspace}
\usepackage[caption=false]{subfig}
\usepackage{float}
\newcommand{\be}{\begin{equation}}
\newcommand{\ee}{\end{equation}}
\newcommand{\bea}{\begin{eqnarray}}
\newcommand{\eea}{\end{eqnarray}}
\newcommand{\void}[1]{}

\newcommand{\summe}[2]{\ensuremath{\sum\limits_{#1}^{#2}}}

\newcommand{\infsumme}[1]{\ensuremath{\sum\limits_{#1}^{\infty}}}

\def\timed{\ensuremath{\displaystyle{\stackrel{\leftrightarrow}{\frac{\partial}{\partial t}}}}}
\def\zeit{\ensuremath{\displaystyle{\frac{\mbox{d}}{\mbox{d}t}}}}
\renewcommand{\i}{\ensuremath{\mathrm{i}}}

\def\N{\ensuremath{\mathbb{N}}\xspace}
\def\C{\ensuremath{\mathbb{C}}\xspace}
\newcommand{\newoperator}[2]{\csdef{#1}{\ensuremath{\hat{#2}}\xspace}%
			     \csdef{#1ad}{\ensuremath{{\hat{#2}}^{\dagger}}\xspace}} 

\newoperator{a}{a}
\newoperator{r}{\rho}
\newoperator{H}{\mathcal{H}}
\newoperator{D}{D}
\newoperator{s}{\sigma}

\newcommand{\HSB}{\ensuremath{\subs{\H}{SB}}\xspace}

\def\laguerre{L}

\def\T{\ensuremath{\mathcal{T}}}
\def\U{\ensuremath{\mathcal{U}}}

\newcommand{\newvariable}[1]{\csdef{#1}{\ensuremath{#1}\xspace}} 	

\newvariable{m}
\newvariable{A}
\newvariable{B}
\newvariable{f}
\newvariable{g}

\newcommand{\newgreek}[2]{\LetLtxMacro{#2}{#1}%
	\renewcommand{#1}{\ensuremath{#2}\xspace}}

\newgreek{\omega}{\oldomega}
\newgreek{\Delta}{\oldDelta}
\newgreek{\alpha}{\oldalpha}
\newgreek{\lambda}{\oldlambda}

\makeatletter
\renewcommand*\env@matrix[1][*\c@MaxMatrixCols c]{%
  \hskip -\arraycolsep
  \let\@ifnextchar\new@ifnextchar
  \array{#1}}
\makeatother

\renewcommand{\Re}[1]{\ensuremath{\operatorname{Re}\left(#1\right)}}

\newcommand{\subs}[2]{\ensuremath{{#1}_{\mbox{\tiny{#2}}}}}
\newcommand{\sups}[2]{\ensuremath{{#1}^{\mbox{\tiny{#2}}}}}

\newcommand{\abs}[1]{\ensuremath{\left|#1\right|}}

\newcommand{\PsiD}{\ensuremath{\sups{\Psi}{D1}}\xspace}
\newcommand{\LD}{\ensuremath{\sups{L}{D1}}\xspace}

\newcommand{\PsiTA}{\ensuremath{\sups{\Psi}{TA}}\xspace}

\newcommand{\HTA}{\ensuremath{\sups{\mathcal{H}}{TA}}\xspace}
\newcommand{\PTA}{\ensuremath{P_z^{\mbox{\tiny{TA}}}}\xspace}

\newcommand{\PsiT}{\ensuremath{\sups{\Psi}{D1}}\xspace}
\newcommand{\PsiTi}{\ensuremath{\sups{\Psi}{D1}_{i}}\xspace}
\newcommand{\PhiT}{\ensuremath{\sups{\Phi}{}}\xspace}
\newcommand{\PhiTi}{\ensuremath{\sups{\Phi}{}_{i}}\xspace}
\newcommand{\PT}{\ensuremath{P_z^{\mbox{\tiny{D1}}}}\xspace}

\newcommand{\PsiBn}{\ensuremath{\sups{\Psi}{B}_{n}}\xspace}

\newcommand{\PF}{\ensuremath{P_z^{\mbox{\tiny{qm}}}}\xspace}
\newcommand{\matlab}{\texttt{matlab }}
\newcommand{\ode}{\texttt{ode15s }}

\begin{document}
\title{Including temperature in a wavefunction description of the dynamics of the quantum Rabi model}
\author{Michael Werther}
\affiliation{Max-Planck-Institut f\"ur Physik Komplexer
Systeme, N\"othnitzer Str. 38, 01187 Dresden, Germany}
\affiliation{Institut f\"ur Theoretische Physik, Technische Universit\"at
Dresden, D-01062 Dresden, Germany}
\author{Frank Grossmann}
\affiliation{Institut f\"ur Theoretische Physik, Technische Universit\"at
Dresden, D-01062 Dresden, Germany}

\date{\today}

\begin{abstract}
We present a wavefunction methodology to account for finite temperature initial conditions
in the quantum Rabi model. The approach is based on the Davydov-Ansatz together with a statistical sampling of the
canonical harmonic oscillator initial density matrix. Equations of motion are gained
from a variational principle and numerical results are compared to those of the thermal Hamiltonian
approach. For a system consisting of a single spin and a single oscillator and for moderate coupling
strength, we compare our new results with full quantum ones as well as with other
Davydov-type results based on alternative sampling/summation strategies. All of these perform better than the ones based on
the thermal Hamiltonian approach. The best agreement is shown by a Boltzmann weighting of individual
eigenstate propagations. Extending this to a bath of many oscillators will, however, be very demanding
numerically. The use of any one of the investigated stochastic sampling approaches will then be favorable.
\end{abstract}
\maketitle

\section{Introduction}

The investigation of the dissipative two-level or spin-boson problem has a long history. An early review with 
a discussion of the ubiquitous appearance of two-level systems in physics as well as of the
solution of their (reduced) dynamics has been given by Leggett and coauthors \cite{Letal87}. 
Two different lines of research have emerged in recent years, depending on the description of the 
dissipative environment. This can either be modeled by harmonic oscillators with a continuous spectral 
density (case I), or by a finite number of oscillators, that could, e.g., result from a discretization 
of a continuous spectral density (case II). The extreme case of just one single oscillator interacting with a 
single spin degree of freedom has a history even longer than that of the spin-boson model and is refered to as 
the quantum Rabi model \cite{Ra36,Ra37,Bra11}. 

In case I, methods of choice for the investigation of the dissipative two-state dynamics are the real-time path integral technique \cite{Wei12},
as well as imaginary time path integral approaches \cite{WRVB09}.
The real-time path integral method allows for exact as well as approximate analytical solutions, and, using
e.g., the quasi-adiabatic path integral (QUAPI) method, also for numerical treatment \cite{NT10},
which then again requires some form of discretization \cite{Mak95}. Another possible numerical way to
determine the reduced density matrix is given by the solution of a stochastic Liouville von Neumann equation \cite{St04}.
In case II, the idea is to treat the dynamics of the composite system, either in some
approximation, or if at all possible, by using a full solution of the underlying 
dynamical equation. In the publications by the Miller group \cite{WTM01,TWM01} 
several tens up to hundreds of bath degrees of freedom have been taken into account in an approximate 
hybrid methodology, whereas a logarithmic discretization of the bath's spectral density has been used
in studies based on the numerical renormalization group \cite{BLTV05}.

In dynamical investigations for finite temperatures, the initial state to be propagated is frequently taken as
a direct product of an initial state of the spin system times a thermal density matrix at the
given temperature of the oscillator bath. One then determines the dynamics of the reduced density matrix 
in case I or the solution of the Liouville-von Neumann equation of the composite system
in case II. For zero temperature and in case II, matters simplify considerably, because
one can solve for the dynamics of the wavefunction, which requires much less memory capacity in a numerical implementation.

Still, the problem of exponential scaling of the numerical effort with increasing system size by taking into
account more and more environmental oscillators is also present in wavefunction
calculations and approximate but accurate enough methods are sought for. Candidates are semiclassical methods \cite{jcp06} or the so-called
Davydov-Ans\"atze  \cite{Da80russ}. In the latter case, the total wavefunction is written as a sum of products of the two spin states
with different (D1-Ansatz), or, even simpler, with the same (D2-Ansatz) coherent harmonic oscillator state.
Equations of motions for the unknown coefficients and coherent state parameters can be gained from a Dirac-Frenkel variational
principle. It has been shown that this method can be put to good use in the spin-boson problem even in the notoriously difficult
case of sub-Ohmic spectral densities \cite{KA13,YDLWZ13}. In addition a way to increase the complexity of the Ansatz and thereby
the accuracy of the solution by using squeezed, instead of coherent states has been pointed out recently \cite{cp16}.

In the following, we intend to answer the question if and how the Davydov D1-Ansatz can be extended to finite temperatures, but 
keeping the simplicity of a description of the dynamics on the wavefunction level. It will be shown that to this
end it is not enough to consider a single wavefunction evolving under a thermalized Hamiltonians as proposed in \cite{Cetal88}.  Similar in spirit
to work by Wang and Thoss \cite{WT06} who used stochastically sampled wavefunctions that are propagated in imaginary as well as in real time, 
one can prescribe a sampling procedure for expectation values of observables at finite temperature using Davydov Ans\"atze \cite{CGVA16,WFCZ17}. Alternatively, 
in the so-called thermo field dynamics, the system Hilbert space is doubled and one can treat the higher dimensional system at temperature zero on the wavefunction level.
This procedure has been put to good use in models of quantum state diffusion \cite{DGS98} as well as in a recent study of
quantum electron vibrational dynamics\cite{BG16}. As a new twist, here, we intend to use a stochastic approach put forth in the calculation of thermal 
rate constants by Matzkies and Manthe \cite{MM99}. Our intention in the following is to present a proof of principle that one can include temperature on the
wavefunction level, i.\ e., that one can mimic the dynamics of a canonical density matrix initial
state in the oscillator Hilbert space by a suitable averaging procedure to be detailed below.
To keep the discussion simple and the numerics easily feasible, in the following, we restrict the discussion to a ``bath''
consisting of just a single harmonic oscillator at a certain temperature $T$. That is, we will study the
dynamics of the quantum Rabi model.

The paper is organized as follows: In Section II, we briefly review the Hamiltonian and the solution of the
corresponding time-dependent Schr\"odinger equation for $T=0$, stressing the importance of the use
of a Dirac-Frenkel variational principle to determine the differential equations for the wavefunction parameters.
In Section III the sampling procedure by Matzkies and Manthe
will be used to arrive at a statistically correct description of temperature by mimicking a canonical initial
density. The resulting equations of motion for the parameters of a Davydov D1 Ansatz with this initial condition are then derived.
An alternative sampling of the $P$-function as well as a Boltzmannized superposition of individual wavefunction propagations
are briefly discussed.
In Section IV, we then compare numerical results of different levels of accuracy with full quantum results
and discuss the problem of singularities in the coupled nonlinear equations of motion in some detail.
In Section V conclusions and an outlook are given.

\section{Quantum Rabi model and Davydov Ansatz for zero temperature}
\label{sec:dav}

The Hamiltonian that will be considered throughout is that of a spin 1/2 system with Hilbert space $\{\ket{+},\ket{-}\}$,
coupled to a harmonic oscillator of frequency $\omega$ and with creation and annihilation operators
$\aad,\a$. This system is governed by the quantum Rabi model \cite{Bra11} (a spin-boson model with a single bosonic mode)
\be
\label{eq:SBham}
\HSB=\frac{\varepsilon}{2}\s_z+V\s_x+\hbar\omega\aad\a+\frac{\lambda}{2}\s_z(\aad+\a)
\ee
and we have suppressed the zero point energy of the oscillator.

The widely used Davydov D1-Ansatz for the solution of the corresponding time-dependent Schr\"odinger equation reads \cite{Da80russ}
\bea
\label{eq:Davy}
\ket{\PsiD(t)} &=& A(t)\ket{+} \D[f(t)]\ket{0}+B(t)\ket{-} \D[g(t)]\ket{0}.
\eea
Here the displacement operator
\be
\D[f(t)]\equiv D_f=\exp\left[f(t) \hat{a}^+-f^{\ast}(t)\hat{a}\right]
\ee
has been introduced. Its action on the ground state $\ket{0}$ of the harmonic oscillator generates a so-called coherent state.

The equations of motion of the time-dependent parameters $A(t),B(t),f(t),g(t)$ can be derived from the Dirac-Frenkel variational principle in its
Lagrangian form \cite{Di30,Fren34,KS81} with the Davydov Dirac-Frenkel Lagrangian 
\be
\label{eq:DDF}
\LD=\bra{\PsiD(t)}
\left(\frac{{\rm i}\hbar}{2}\frac{\stackrel{\leftrightarrow}{\partial}}{\partial t}-\HSB\right)
\ket{\PsiD(t)}.
\ee
Using the orthogonality of the spin states and the time derivative of the displacement operator, we get
\bea
\bra{\PsiD(t)}\left(\frac{{\rm i}\hbar}{2}\frac{\stackrel{\leftrightarrow}{\partial}}{\partial t}\right)\ket{\PsiD(t)}
&=&\frac{\rm i\hbar}{2}\left\{A^\ast\dot{A}-A\dot{A}^\ast+B^\ast\dot{B}-B\dot{B}^\ast\right .
\nonumber
\\
&&\left .+|A|^2(f^\ast\dot{f}-f\dot{f}^\ast)+|B|^2(g^\ast\dot{g}-g\dot{g}^\ast)\right\}
\label{eq:tder}
\eea
for the time derivative and
\bea
\bra{\PsiD(t)}\HSB\ket{\PsiD(t)}
&=&\frac{\varepsilon}{2}\left(|A|^2-|B|^2\right)
+V\left(A^\ast B{\rm e}^{f^\ast g}+A B^\ast{\rm e}^{fg^\ast}\right)
{\rm e}^{-\frac{1}{2}(|f|^2+|g|^2)}
\nonumber
\\
&&+\hbar\omega\left(|A|^2|f|^2+|B|^2|g|^2\right)
\nonumber
\\
&&+\frac{\lambda}{2}\left\{|A|^2(f+f^\ast)-|B|^2(g+g^\ast)\right\}
\label{eq:hamD}
\eea
for the Hamiltonian part.

The Eulerian equations of motion
\be
\frac{\rm d}{{\rm d}t}\frac{\partial \LD}{\partial \dot{u}_i^\ast}-\frac{\partial \LD}{\partial u_i^\ast}=0,
\label{eq:euler}
\ee
for the wavefunction parameters $u_i\in\{A,B,f,g\}$ are given by \cite{WDLZ13,cp16}
\bea
0&=&{\rm i}\hbar\dot{A}+ \frac{{\rm i}\hbar}{2}A(\dot{f}f^\ast-f\dot{f}^\ast)-V B{\rm e}^{f^\ast~g}
{\rm e}^{-\frac{1}{2}(|f|^2+|g|^2)}-\hbar\omega A |f|^2-\frac{\lambda}{2}A(f+f^\ast)-\frac{\varepsilon}{2}A
\label{eq:davy1}
\\
0&=&{\rm i}\hbar\dot{B}+ \frac{{\rm i}\hbar}{2}B(\dot{g}g^\ast-g\dot{g}^\ast)-V A{\rm e}^{g^\ast~f}
{\rm e}^{-\frac{1}{2}(|f|^2+|g|^2)}-\hbar\omega B |g|^2+\frac{\lambda}{2}B(g+g^\ast)+\frac{\varepsilon}{2}B
\label{eq:davy2}
\\
0&=&i\hbar A\dot{f}-V B~(g-f){\rm e}^{f^\ast g}{\rm e}^{-\frac{1}{2}(|f|^2+|g|^2)}-\hbar\omega Af-\frac{\lambda}{2} A
\label{eq:davy3}
\\
0&=&i\hbar B\dot{g}-V A~(f-g){\rm e}^{g^\ast f}{\rm e}^{-\frac{1}{2}(|f|^2+|g|^2)}-\hbar\omega Bg+\frac{\lambda}{2} B.
\label{eq:davy4}
\eea
These are 4 implicit ordinary differential equations (ODE) of complex variables, i.e., in total these
are 8 real ODEs. Division of \eqref{eq:davy3} by $A$, and replacing the second term of \eqref{eq:davy1} by the result, yields explicit equations, which can be solved numerically. Nevertheless it is important to note that for $A=0$ (resp. $B=0$), which occurs at several times that depend on the constellation of the parameters  $\varepsilon$, $V$, $\omega$, $\lambda$ in \eqref{eq:SBham}, the resulting equations for $\dot f$ (resp. $\dot g$) have singularities. To be more precise: to derive \eqref{eq:davy3} it has been divided by a factor $A^\ast$, so the original result reduces to $0=0$ for $A=0$. This takes into account the fact that for $A=0$ in \eqref{eq:Davy}, $f$ can be chosen arbitrarily. We will come back to this later in \ref{sec:Num}. \\     
The initial conditions 
\be
\label{eq:ic}
A(0)=1,B(0)=0,f(0)=0,g(0)=0
\ee
are frequently used in numerical studies, representing a wavefunction in the up state of the spin system and in the
ground state of the harmonic degree of freedom.

An extension of the complexity of the Davydov-Ansatz by using squeezed states has been given in
\cite{cp16}, where also some remarks regarding the numerical solution of the (extended) equations of motion can be found.  
Furthermore, including more environmental degrees of freedom is straightforward. The corresponding equations of motion
have been detailed in \cite{WDLZ13}.

\section{Davydov Ansatz for propagation of the canonical density matrix}

In this section we now turn to the question how the canonical density matrix can be propagated on the level of wavefunctions. 
The reason why we are interested in finding an answer to this question is twofold. 
Firstly, working with wavefunctions minimizes the storage requirements in a numerical implementation of the dynamics. 
Secondly, and even more important in the present context is the fact that the Davydov-Ansatz and the Dirac-Frenkel variational principle 
can only be formulated on the wavefunction level. No analogous formulation leading to suitable working equations is known on the density matrix 
level of description.

\subsection{Davydov Ansatz with thermal averaging} \label{sec:thermAver}

The thermal stability of solitonic solutions of energy transport through proteins within the Davydov Ansatz was a hot topic in the 80s and
early 90s of the last century. Analogous to the work of Cruzeiro et al.\ \cite{Cetal88} and by F\"orner \cite{For92} for exciton-phonon
models, one could be tempted to include temperature effects in a spin-boson model by using a generalized Davydov Ansatz  
\bea
\label{eq:TA}
\ket{\PsiTA_{n}(t)} &=& A(t)\ket{+} \D[f(t)]\ket{n}+B(t)\ket{-} \D[g(t)]\ket{n},
\eea
with the normalized excited states
\be
\ket{n} =\frac{1}{\sqrt{n!}}(\aad)^n\ket{0}
\ee
and a thermally averaged Hamiltonian
\be
\HTA=\sum_{n=0}^\infty\rho_n \HTA_{nn}
\ee
with
\be
\rho_{n}=\frac{\bra{n}{\rm e}^{-\beta\hbar\omega \hat{a}^+\hat{a}}\ket{n}}
{\infsumme{n=0}\bra{n}{\rm e}^{-\beta\hbar\omega \hat{a}^+\hat{a}}\ket{n}}
=\frac{{\rm e}^{-\beta\hbar\omega n}}{Q(\beta)},
\ee
where 
\be
\label{eq:part}
Q(\beta)=\infsumme{n=0}{\rm e}^{-\beta E_{n}}=\left(1-{\rm e}^{-\beta\hbar\omega}\right)^{-1}
\ee
is the canonical partition function, 
\be
E_n=n\hbar\omega
\ee
are the harmonic oscillator eigenvalues (without the groundstate energy), $\beta=1/\subs{k}{B}T$ is proportional to the inverse temperature 
with Boltzmann constant $\subs{k}{B}$, and 
\be
\HTA_{nn}=\bra{\PsiTA_n(t)}\HSB\ket{\PsiTA_n(t)}
\ee
is the diagonal matrix element of the Hamiltonian between the Davydov wavefunctions with excited harmonic states.

In the appendix of \cite{pccp17} it has been shown numerically that the equations of motion that arise
from the thermally averaged Hamiltonian do not account for the correct dynamics, however. This can have three reasons. 
Either it is due to the limitation of the Davydov-Ansatz, or it is due to numerical issues which could come into play because of singular equations for $\dot f$ (resp $\dot g$), 
or it is due to an incorrect inclusion of temperature. The second reason will be treated in more detail in Section \ref{sec:Num}. The third reason can be further elucidated by allowing 
for a stochastic element in the wavefunction calculation, thereby mimicking the exact density matrix equation that should be used in the presence of temperature.

\subsection{Stochastic wavefunctions for the quantum Rabi model}\label{ssec:eofm}

\begin{figure}
\includegraphics[scale=0.3,trim= 0cm 0cm 0cm 0cm,clip=True]{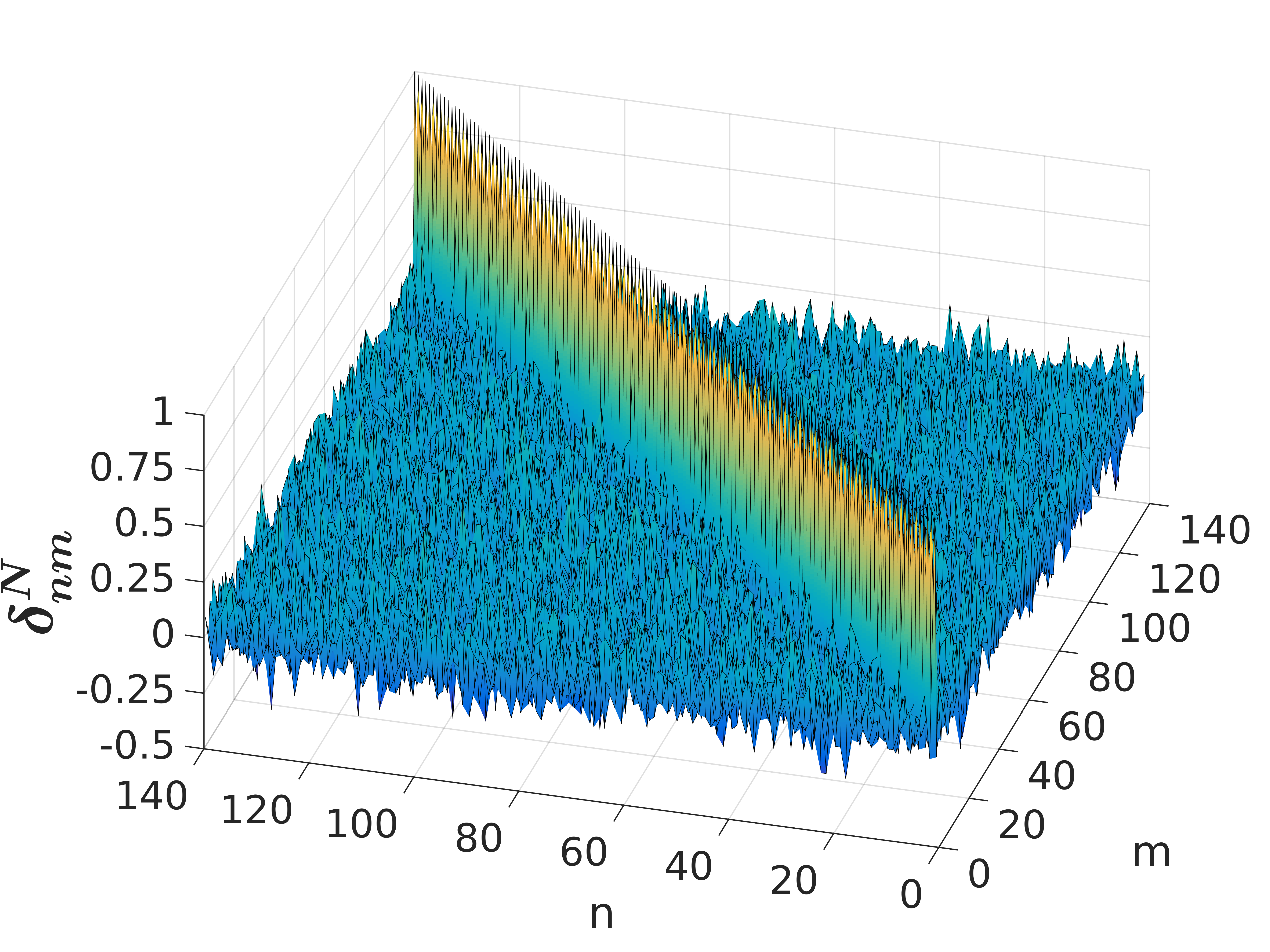}
\caption{Realization of the Kronecker's $\delta$ via a stochastic sampling according to
(\ref{eq:delta}) using $N=100$.}
\label{fig:kron}
\end{figure}

For a correct description of temperature,
let us start with a Boltzmannized superposition of eigenstates of the harmonic oscillator
\be
\label{eq:Phi}
\ket{\PhiTi}=\frac{1}{\sqrt{Q(\beta)}}\infsumme{n=0}(-1)^{\alpha_{{n}}^i}{\rm e}^{-\frac{\beta E_{n}}{2}}\ket{n},
\ee
where the $\alpha_n^i$ are random integers that can be either plus or minus one.
With a statistical average 
over many realizations of the random numbers the canonical density matrix can be generated as
\bea
\hat{\rho}&=&\frac{1}{N}\sum_{i=1}^N\ket{\PhiTi}\bra{\PhiTi}
\nonumber
\\
&=&\frac{1}{Q(\beta)}\sum_{{n},{m}}\frac{1}{N}\sum_{i=1}^N(-1)^{\alpha_{n}^i+\alpha_{m}^i}
{\rm e}^{-\frac{\beta}{2}\left(E_{n}+E_{m}\right)}\ket{n}\bra{m}
\nonumber
\\
\label{eq:rho}
&=&\frac{1}{Q(\beta)}\sum_{{n}}{\rm e}^{-\beta E_{n}}\ket{n}\bra{n},
\eea
where the last equation strictly only holds for $N\to\infty$ because in this limit
\be
\label{eq:delta}
\frac{1}{N}\sum_{i=1}^N(-1)^{\alpha_{n}^i+\alpha_{m}^i}=\delta_{{n}{m}}.
\ee
A realization of the Kronecker-$\delta$ for a finite number $N=100$ is shown in Figure \ref{fig:kron}. 
Therefore, in a numerical implementation a few tens of realizations may be enough to
accurately mimic a canonical density matrix by this procedure. 
In passing we note that not only ${\rm Tr}\hat\rho=1$, but also on the level of individual $i$ 
\[\braket{\PhiTi|\PhiTi}=1\]
holds.

The calculation of expectation values of operators using the indicated averaging procedure has been discussed
by Matzkies and Manthe \cite{MM99}, based on earlier work by Jeffrey and Smith \cite{JS97}. It has been
shown that if the operator in question has the same eigenstates as the one used in the
expansion of the wavefunction then a single realization is enough, whereas in the general case the convergence
of the error to zero is given by $1/\sqrt{N}$.


We now turn to the quantum Rabi model (\ref{eq:SBham}) with $\varepsilon=0$ and employ the Ansatz 
\be
\ket{\PsiTi(t)} = A_i(t)\ket +\D[f_i(t)]\ket{\PhiTi}+ B_i(t)\ket -\D[g_i(t)]\ket{\PhiTi},
\ee
with $\ket{\PhiTi}$ defined in (\ref{eq:Phi}), for the solution of the quantum dynamical problem with a 
thermal initial condition in the oscillator Hilbert space.

The equations of motion for the coefficients can again be derived from the Dirac-Frenkel variational principle. 
In the sequel we calculate the Lagrangian density and, for clarity of notation, we temporarily suppress the index $i$. For the time-derivative, we get
\bea
\bra{\PsiT}\timed\ket{\PsiT} &=& \dot AA^\ast-A\dot A^\ast+\dot BB^\ast-B\dot B^\ast
		+|A|^2\left[\dot ff^\ast-f\dot f^\ast+2\left(\dot f-\dot f^\ast\right)\T_0^1\right]  \notag \\ & &
		+|B|^2\left[\dot gg^\ast-g\dot g^\ast+2\left(\dot g-\dot g^\ast\right)\T_0^1\right],\qquad
\eea
where we define for integer numbers $r,s\in\N$ time-independent quantities:
\bea
\T_s^r:&=&\bra{\PhiT}\left(\aad\right)^r\left(\a\right)^s\ket{\PhiT} \notag \\
&=& \frac 1{Q\left(\beta\right)}\infsumme{n=0}(-1)^{\alpha_{n+s}+\alpha_{n+r}}{\rm e}^{-\frac{\beta}{2}\left(E_{n+s}+E_{n+r}\right)}\frac{\sqrt{(n+s)!(n+r)!}}{n!}
\eea
and
\bea
\U_r^s:&=&\bra{\PhiT}\left(\a\right)^r\left(\aad\right)^s\ket{\PhiT} \notag \\
&=& \frac 1{Q\left(\beta\right)}\infsumme{n=\max\{r,s\}}(-1)^{\alpha_{n-s}+\alpha_{n-r}}{\rm e}^{-\frac{\beta}{2}\left(E_{n-s}+E_{n-r}\right)}\frac{n!}{\sqrt{(n-s)!(n-r)!}}.
\eea
Tedious but straightforward calculations show that
\bea
\T^s_s = \frac{s!}{\left({\rm e}^{\beta\hbar\omega}-1\right)^s}, \qquad \U^s_s = \frac{s!}{\left(1-{\rm e}^{-\beta\hbar\omega}\right)^s}, \qquad \T^s_r = \T^r_s, \qquad \U^s_r = \U^r_s, \qquad \U^s_0 = \T^s_0
\eea
holds.
For the Hamiltonian part of the Lagrangian, we obtain
\bea
\bra{\PsiT}\H_{SB}\ket{\PsiT} &=& \frac{\lambda}{2}\left[|A|^2\left(f+f^\ast+2\T_0^1\right)-|B|^2\left(g+g^\ast+2\T_0^1\right)\right]\notag \\
& & +\hbar\omega\left[|A|^2\left(|f|^2+(f^\ast+f)\T_0^1+\T_1^1\right)+|B|^2\left(|g|^2+(g^\ast+g)\T_0^1+\T_1^1\right)\right] \notag \\
& & +V\left[A^\ast B\bra{\PhiT}\Dad_f\D_g\ket{\PhiT}+B^\ast A\bra{\PhiT}\Dad_g\D_f\ket{\PhiT}\right].
\eea
The equations of motion are again the Eulerian equations of motion, yielding
\bea
0 &=& \i\hbar\dot A + \frac{\i\hbar}{2}A\left[\dot ff^\ast-f\dot f^\ast+2\left(\dot f-\dot f^\ast\right)\T_0^1\right]-
V B\bra{\PhiT}\Dad_f\D_g\ket{\PhiT} \notag \\ & & -\hbar\omega A \left(|f|^2+(f+f^\ast)\T_0^1+\T_1^1\right)  -\frac\lambda 2 A \left(f+f^\ast+2\T_0^1\right)
\label{eq:davystoch1}
\eea
for $A(t)$ and a similar equation for $B(t)$, with a sign change in the last term.

We note that due to equation \eqref{eq:davystoch1} 
\bea \zeit |A|^2 &=& 2\Re{A^\ast\dot A} = 2\Re{-V\frac{\i}{\hbar} A^\ast B\bra{\PhiT}\Dad_f\D_g\ket{\PhiT}} \notag \\
&=& -V\frac{\i}{\hbar}\left(A^\ast B\bra{\PhiT}\Dad_f\D_g\ket{\PhiT}-A B^\ast \bra{\PhiT}\Dad_g\D_f\ket{\PhiT}\right).\qquad
\label{eq:davystoch1a} 
\eea
Furthermore we use
\bea
\Dad_f\D_g = \D_{-f}\D_g = {\rm e}^{\frac 12\left(f^\ast g-fg^\ast\right)}\D_{g-f} = {\rm e}^{-\frac 12\left(|f|^2+|g|^2\right)}{\rm e}^{f^\ast g}{\rm e}^{(g-f)\aad}{\rm e}^{-\left(g^\ast-f^\ast\right)\a},
\label{eq:displace}
\eea
from which follows that 
\bea
\frac{\partial}{\partial f^\ast}\Dad_f\D_g &=& \Dad_f\D_g\left(-\frac 12 f+g+\a\right), \notag \\
\frac{\partial}{\partial f^\ast}\Dad_g\D_f &=& \Dad_g\D_f\left(-\frac 12 f-\a\right). \notag
\eea
Thus we obtain
\bea
0 &=& \i\hbar |A|^2\dot f-\frac\lambda 2 |A|^2-\hbar\omega |A|^2\left(f+\T_0^1\right) \notag \\
 && -V \left\{A^\ast B\left[\left(g-f-\T_0^1\right)\bra{\PhiT}\Dad_f\D_g\ket{\PhiT}+\bra{\PhiT}\Dad_f\D_g\a\ket{\PhiT}\right]\right. \notag \\
 && \left. +AB^\ast \left[\T_0^1 \bra{\PhiT}\Dad_g\D_f\ket{\PhiT} -\bra{\PhiT}\Dad_g\D_f\a\ket{\PhiT}\right]\right\}.
\label{eq:fdot}
\eea
Equations (\ref{eq:davystoch1},\ref{eq:fdot}) and the corresponding ones for $B(t)$ and $g(t)$, which can be
gained from the previous ones by interchanging $A(t)$ and $B(t)$ as well as $f(t)$ and $g(t)$ with an additional sign change in the term proportional to $\lambda$
are our working equations. 

To solve them numerically, explicit equations can be obtained by division of \eqref{eq:fdot} by $\abs{A}^2$ and insertion of the result into \eqref{eq:davystoch1}. 
Expectation values of the form $\bra{\Phi}\Dad_f\D_g\ket{\Phi}$ and $\bra{\Phi}\Dad_f\D_g\a\ket{\Phi}$ cannot be expressed analytically due to the random signs $(-1)^{\alpha_n^i}$ in $\ket{\Phi_i}$. 
Furthermore, they are time-dependent due to the time-dependency of $f$ and $g$, so they have to be calculated in each integration step. To do this efficiently, we define for $s\in\N$ time-independent 
vectors $v_s$ component-wise by
\be
\left(v_s\right)_n:=\frac{{\rm e}^{-\frac{\beta}{2}E_{n+s}}}{\sqrt Q}\sqrt{\frac{(n+s)!}{n!}} (-1)^{\alpha_{n+s}}.
\ee
The matrix elements ($m,n\in\N$, $z\in\C$)
\bea \label{eq:overlapmn}
M_{mn}(z):=\bra m e	^{z\a}e^{-z^\ast\aad}\ket n = \summe{k=0}{\min\{m,n\}}\frac{z^{m-k}\left(-z^\ast\right)^{n-k}}{(m-k)!(n-k)!k!}\sqrt{m!n!},
\eea
can be rewritten by introducing the associated Laguerre polynomials ($n,k\in\N$, $x\in\C$) 
\be
\laguerre_n^k(x)=\summe{i=0}{n}\frac{1}{i!} \binom{k+n}{n-i}(-x)^i, \label{eq:lagu}
\ee 
 as
\bea
M_{mn}(z)= \begin{cases} \left(-z^\ast\right)^{n-m}\sqrt{\frac{m!}{n!}}\,\laguerre_{m}^{n-m}\left(\abs{z}^2\right) & \text{for } m\leq n \\ z^{m-n}\sqrt{\frac{n!}{m!}}\,\laguerre_{n}^{m-n}\left(\abs{z}^2\right) & \text{for } n\leq m\end{cases}
\eea
Thus by using \eqref{eq:displace}
\be
\label{eq:stoch_expval}
\bra{\Phi}\Dad_f\D_g\ket{\Phi}={\rm e}^{-\frac 12\left(|f|^2+|g|^2\right)}{\rm e}^{f^\ast g} \vec{v}_0\cdot {\bf M}(g-f) \vec{v}_0. 
\ee
A similar expression with shifted index can be obtained for $\bra\Phi\Dad_f\D_g\a\ket\Phi$. Furthermore, $\T_0^1=\vec{v}_0\cdot \vec{v}_1$. Despite these simplifications, evaluation of the right-hand-side 
of the explicit system of differential equations is - especially compared to the method employed in Section \ref{sec:Bol}- very expensive.


We note first that the Dirac-Frenkel variational principle is only valid for $\braket{\PsiT|\PsiT}=1$, i.e.
\bea
|A|^2+|B|^2=1
\eea
is required to hold for all times, which indeed is a direct consequence of equation \eqref{eq:davystoch1a} if we require it to hold for $t=0$.
\void{
We have the following working equations:
\bea
0 &=& \i\hbar\dot A + \frac{\i\hbar}{2}A\left[\dot ff^\ast-f\dot f^\ast+2\left(\dot f-\dot f^\ast\right)\T_0^1\right]-V B\bra{\PhiT}\Dad_f\D_g\ket{\PhiT} \notag \\ & & -\hbar\omega A \left(|f|^2+(f+f^\ast)\T_0^1+\T_1^1\right)  -\frac\lambda 2 A \left(f+f^\ast+2\T_0^1\right) \\
0 &=& \i\hbar\dot B + \frac{\i\hbar}{2}B\left[\dot gg^\ast-g\dot g^\ast+2\left(\dot g-\dot g^\ast\right)\T_0^1\right]-V A\bra{\PhiT}\Dad_g\D_f\ket{\PhiT} \notag \\ & & -\hbar\omega B \left(|g|^2+(g+g^\ast)\T_0^1+\T_1^1\right)  +\frac\lambda 2 B \left(g+g^\ast+2\T_0^1\right) \\
0 &=& \i\hbar |A|^2\dot f-\frac\lambda 2 |A|^2-\hbar\omega |A|^2\left(f+\T_0^1\right) \notag \\
 && -V \left\{A^\ast B\left[\left(g-f-\T_0^1\right)\bra{\PhiT}\Dad_f\D_g\ket{\PhiT}+\bra{\PhiT}\Dad_f\D_g\a\ket{\PhiT}\right]\right. \notag \\
 && \left. +AB^\ast \left[\T_0^1 \bra{\PhiT}\Dad_g\D_f\ket{\PhiT} -\bra{\PhiT}\Dad_g\D_f\a\ket{\PhiT}\right]\right\} \\
 0 &=& \i\hbar |B|^2\dot g+\frac\lambda 2 |B|^2-\hbar\omega |B|^2\left(g+\T_0^1\right) \notag \\
 && -V \left\{B^\ast A\left[\left(f-g-\T_0^1\right)\bra{\PhiT}\Dad_g\D_f\ket{\PhiT}+\bra{\PhiT}\Dad_g\D_f\a\ket{\PhiT}\right]\right. \notag \\
 && \left. +A^\ast B \left[\T_0^1 \bra{\PhiT}\Dad_f\D_g\ket{\PhiT} -\bra{\PhiT}\Dad_f\D_g\a\ket{\PhiT}\right]\right\}
\eea
}

Second, for temperature $T=0$ we have $\ket{\PhiT}=\ket{\PhiTi}=(-1)^{\alpha_0^i}\ket 0$ and hence $\T_0^1=\T_1^1=0$ and 
$\bra{\PhiT}\Dad_f\D_g\ket{\PhiT}={\rm e}^{-\frac 12\left(|f|^2+|g|^2\right)}{\rm e}^{f^\ast g}$ (which can be seen from equation \eqref{eq:displace}) as well as 
$\bra{\PhiT}\Dad_f\D_g\a\ket{\PhiT}=0$, so that the equations obtained reduce for $T=0$ to the ones in Section \ref{sec:dav}, as to be expected. 

Third, upon reintroducing the index $i$ and assuming the parameters $A_i, B_i,f_i, g_i$ not to depend on $i$, the above equations of motion 
(\ref{eq:davystoch1},\ref{eq:fdot}) reduce - by averaging the equations over $N$ and using property \eqref{eq:delta} - to the ones obtained by thermal averaging \cite{pccp17}. As shown in that reference, this does 
not yield the correct dynamics, however.


\subsection{Sampling of $P$-function}\label{sec:PFun}

An alternative approach to correct density matrix dynamics based on a wavefunction approach has been proposed in \cite{WFCZ17}.
The thermal density matrix of the harmonic oscillator in (\ref{eq:rho}) can be expressed in a coherent state basis
$\{\ket{\alpha}\}$ with the help of the so-called $P$-function via
\be
\hat{\rho}=\int{\rm d}^2\alpha P_\beta(\alpha,\alpha^\ast)\ket{\alpha}\bra{\alpha},
\ee
with $\alpha=x+{\rm i}p$ and ${\rm d}^2\alpha={\rm d}x{\rm d}p$ and where \cite{GaZo}
\be
P_\beta(\alpha,\alpha^\ast)=\frac{1}{\pi}\left({\rm e}^{\beta\hbar\omega}-1\right)\exp\left\{-|\alpha|^2\left({\rm e}^{\beta\hbar\omega}-1\right)\right\}.
\ee
The importance sampling of the phase space centers of the bosonic degree of freedom with the weighting function given has been put to good use
in a recent study of the Davydov-Ansatz \cite{WFCZ17}. There the original $T=0$ equations, given in Section \ref{sec:dav}, have been used. 
The initial conditions $f(0)=g(0)=\alpha$ of the harmonic oscillator are not zero, however, but are drawn from the distribution given by the $P$-function. 

This is in contrast to our new approach based on the sampling strategy a la Matzkies and Manthe, where the initial conditions of the oscillator are still 
zero but we use modified, temperature dependent equations.

\subsection{Boltzmann average of individual Davydov wavefunctions} \label{sec:Bol}

In the section on numerical results below, we also compare to the case of direct thermal (Boltzmann) averaging
according to the energy domain representation of the thermal density operator
\bea
\subs{\r}{B}=\frac{1}{Q(\beta)}\infsumme{n=0}{\rm e}^{-\beta E_{n}}\ket{n}\bra{n}
\eea
of the harmonic oscillator. For the quantum Rabi model, the Davydov-Ansatz is then
\bea
\ket{\PsiBn(t)} = A_n(t)\ket{+} \D[f_n(t)]\ket{n}+B_n(t)\ket{-} \D[g_n(t)]\ket{n}
\eea
and the density operator is given by
\be
\label{eq:rhoBol}
\r(t)=\frac{1}{Q(\beta)}\infsumme{n=0}{\rm e}^{-\beta E_{n}}\ket{\PsiBn(t)}\bra{\PsiBn(t)}.
\ee
The Davydov Lagrangian has the same time-derivative part as given in the $T=0$ case in Eq. (\ref{eq:tder}).
The corresponding Hamiltonian part is (suppressing the index $n$ for the parameters)
\bea
\braket{\PsiBn(t)|\subs{\H}{SB}|\PsiBn(t)} 
&=& V{\rm e}^{-\frac 12\left(\abs{f}^2+\abs{g}^2\right)}\laguerre_n\left(\abs{g-f}^2\right) 
\left[A^\ast B{\rm e}^{f^\ast g}+AB^\ast {\rm e}^{fg^\ast}\right]\notag \\
& & +\hbar\omega\left[\abs{A}^2\left(\abs{f}^2+n\right)+\abs{B}^2\left(\abs{g}^2+n\right)\right] \notag \\
& & + \frac{\lambda}{2}\left[\abs{A}^2\left(f+f^\ast\right)-\abs{B}^2\left(g+g^\ast\right)\right].
\eea
From the complete Lagrangian, the equations of motion
\bea
0 &=& \i\hbar\dot{A} + \frac{\i\hbar}{2}A\left(\dot ff^\ast-f\dot f^\ast\right)
			-VB{\rm e}^{-\frac 12\left(\abs{f}^2+\abs{g}^2\right)}{\rm e}^{f^\ast g}\laguerre_n\left(\abs{f-g}^2\right) \notag \\
		& & -\hbar\omega A\left(\abs{f}^2+n\right)-\frac{\lambda}{2}A\left(f+f^\ast\right) \label{eq:Bol1} \\
0 &=& \i\hbar\dot{B} + \frac{\i\hbar}{2}B\left(\dot gg^\ast-g\dot g^\ast\right)
			-VA{\rm e}^{-\frac 12\left(\abs{f}^2+\abs{g}^2\right)}{\rm e}^{g^\ast f}\laguerre_n\left(\abs{f-g}^2\right) \notag \\
		& & -\hbar\omega B\left(\abs{g}^2+n\right)+\frac{\lambda}{2}B\left(g+g^\ast\right) \label{eq:Bol2} \\	
0 &=& \i\hbar\abs{A}^2\dot f -V{\rm e}^{-\frac 12\left(\abs{f}^2+\abs{g}^2\right)}(g-f)\left\{A^\ast B{\rm e}^{f^\ast g}
\left[\laguerre_n\left(\abs{f-g}^2\right)-\laguerre_n'\left(\abs{f-g}^2\right)\right]\right. \notag \\ & & \left.\hphantom{\i\hbar\abs{A}^2\dot f 
-V{\rm e}^{-\frac 12\left(\abs{f}^2+\abs{g}^2\right)}(g-f)\left\{\right.}-AB^\ast {\rm e}^{fg^\ast}\laguerre'_n\left(\abs{f-g}^2\right)\right\} \notag \\
 & & -\hbar\omega\abs{A}^2f-\frac{\lambda}{2}\abs{A}^2 \label{eq:Bol3} \\
0 &=& \i\hbar\abs{B}^2\dot g -V{\rm e}^{-\frac 12\left(\abs{f}^2+\abs{g}^2\right)}(f-g)\left\{AB^\ast {\rm e}^{fg^\ast}
\left[\laguerre_n\left(\abs{f-g}^2\right)-\laguerre_n'\left(\abs{f-g}^2\right)\right]\right. \notag \\ & & \left.\hphantom{\i\hbar\abs{B}^2\dot g 
-V{\rm e}^{-\frac 12\left(\abs{f}^2+\abs{g}^2\right)}(f-g)\left\{\right.}-A^\ast B {\rm e}^{f^\ast g}\laguerre'_n\left(\abs{f-g}^2\right)\right\} \notag \\
 & & -\hbar\omega\abs{B}^2g+\frac{\lambda}{2}\abs{B}^2 \label{eq:Bol4}	
\eea
emerge, which are of much higher complexity than in the case $T=0$ due to the appearance of the Laguerre polynomials (and their derivatives). 
These are defined according to \eqref{eq:lagu} by 
\be
\laguerre_n(x)=L_n^0(x). 
\ee
Firstly, we note that for $n=0$ the equations of motion reduce to the ones given in Section \ref{sec:dav}, which corresponds to the case $T=0$. 
Secondly, for nonzero temperature, the infinite sum in \eqref{eq:rhoBol} 
can for numerical purposes be truncated at $N_T$ depending on temperature $T$.
Thirdly, again upon reintroducing the index $n$ and assuming the parameters $A_n, B_n,f_n, g_n$ not to depend on $n$ (as in Eq.\ (\ref{eq:TA})), 
the above equations of motion reduce - 
by averaging the equations according to the Boltzmann weights in \eqref{eq:rhoBol} and using the generating function of the Laguerre polynomials - to the ones 
obtained by thermal averaging \cite{pccp17}.
We stress that this thermal averaging of the equations {\it before propagation} leads to very different numerical results (see below) as the propagation of 
{\it individual} wavefunctions and Boltzmannizing only at the very end!

\subsection{Expectation values}

With the different approaches to the dynamics presented above, expectation values can be calculated. 
In the following, we focus on the expectation value of the $z$ component of the vector of the Pauli spin matrices, 
also denoted as population difference, representing the (damped) dynamics of the system which is coupled to the bath harmonic oscillator. 
For the thermal averaging it is given by
\be
\PTA(t)=\langle \hat{\sigma}_z\rangle(t)=\infsumme{n=0}\rho_n
\bra{\PsiTA_{n}(t)}\hat{\sigma}_z\ket{\PsiTA_{n}(t)}=|A(t)|^2-|B(t)|^2,
\ee
which depends on temperature through the dependence on temperature of $A(t)$ and $B(t)$.

For the new wave-function approach presented herein, based on stochastic sampling of the thermal density operator,
the population difference is given by
\bea
\PT(t) = \frac 1N\summe{i=1}{N}\left[|A_i(t)|^2-|B_i(t)|^2\right],
\eea
where the dependency on temperature is again given through $A_i$ and $B_i$.

In the case of sampling of initial conditions according to the $P$ function,
we get 
\bea
P_z^{P}(t)=\int{\rm d}^2\alpha P_\beta(\alpha,\alpha^\ast)\left[|A_\alpha(t)|^2-|B_\alpha(t)|^2\right],
\eea
where the $A$ and $B$ coefficients depend on the temperature dependent initial conditions for the bath variables, as 
indicated by the corresponding index.

For the Boltzmann-averaged case, the observable is calculated from
\be
P_z^{\rm B}(t)=\sum_{n=0}^{N_T}\frac{{\rm e}^{-\beta E_n}}{Q}\left[|A_n(t)|^2-|B_n(t)|^2\right],
\ee
where temperature appears explicitly in the exponent.

\section{Comparison of numerical results} \label{sec:Num}

In all the results to be presented below, we choose  
\bea
\r(0)=\ket +\bra +\frac 1{Q(\beta)}\left(\infsumme{n=0}{\rm e}^{-\beta E_n}\ket n\bra n\right)
\eea
as the initial density. This is a direct product of a pure initial system state and the canonical density matrix of the bath oscillator.

A fully quantum and (which is most important for numerical purposes) quickly converging solution of the Rabi model in the case of description 
of the bath by only one oscillator can be obtained by propagating the initial states $|\phi_n(0)\rangle=\ket n\ket +$ for all $n$ by the respective Hamiltonian
under the full time-dependent Schr\"odinger equation. 
If we denote the resulting state by $\ket{\phi_n(t)}$, then with the above initial condition the propagated density is given by
\bea
\label{eq:rhofull}
\r(t)=\frac{1}{Q(\beta)}\infsumme{n=0}{\rm e}^{-\beta E_n}\ket{\phi_n(t)}\bra{\phi_n(t)}.
\eea
This result also contains a Boltzmann average but does not contain any approximation because we do not use the Davydov-Ansatz.
Since $\ket{\phi_n(t)}$ stays normalized for all $t$, this indeed is converging fast, due to the exponential factors. 
By expanding in eigenstates in both Hilbert space dimensions,
\bea
\ket{\phi_n(t)}=\infsumme{j=0}\left(c_{nj}^{(+)}(t)\ket + + c_{nj}^{(-)}(t)\ket -\right)\ket j,
\eea
the population difference becomes
\bea
\PF(t)&=&\frac{1}{Q(\beta)}{\rm Tr}\left(\infsumme{n=0} {\rm e}^{-\beta E_n}\hat{\sigma}_z\ket{\phi_n(t)}\bra{\phi_n(t)}\right)
\nonumber
\\
&=&\frac{1}{Q(\beta)}\infsumme{j=0}\infsumme{n=0}{\rm e}^{-\beta E_n}\left(|c_{nj}^{(+)}(t)|^2-|c_{nj}^{(-)}(t)|^2\right).
\eea
The two sums will be truncated for numerical purposes.
\begin{figure}
\includegraphics[scale=0.3,trim= 0cm 0cm 0cm 0cm,clip=True]{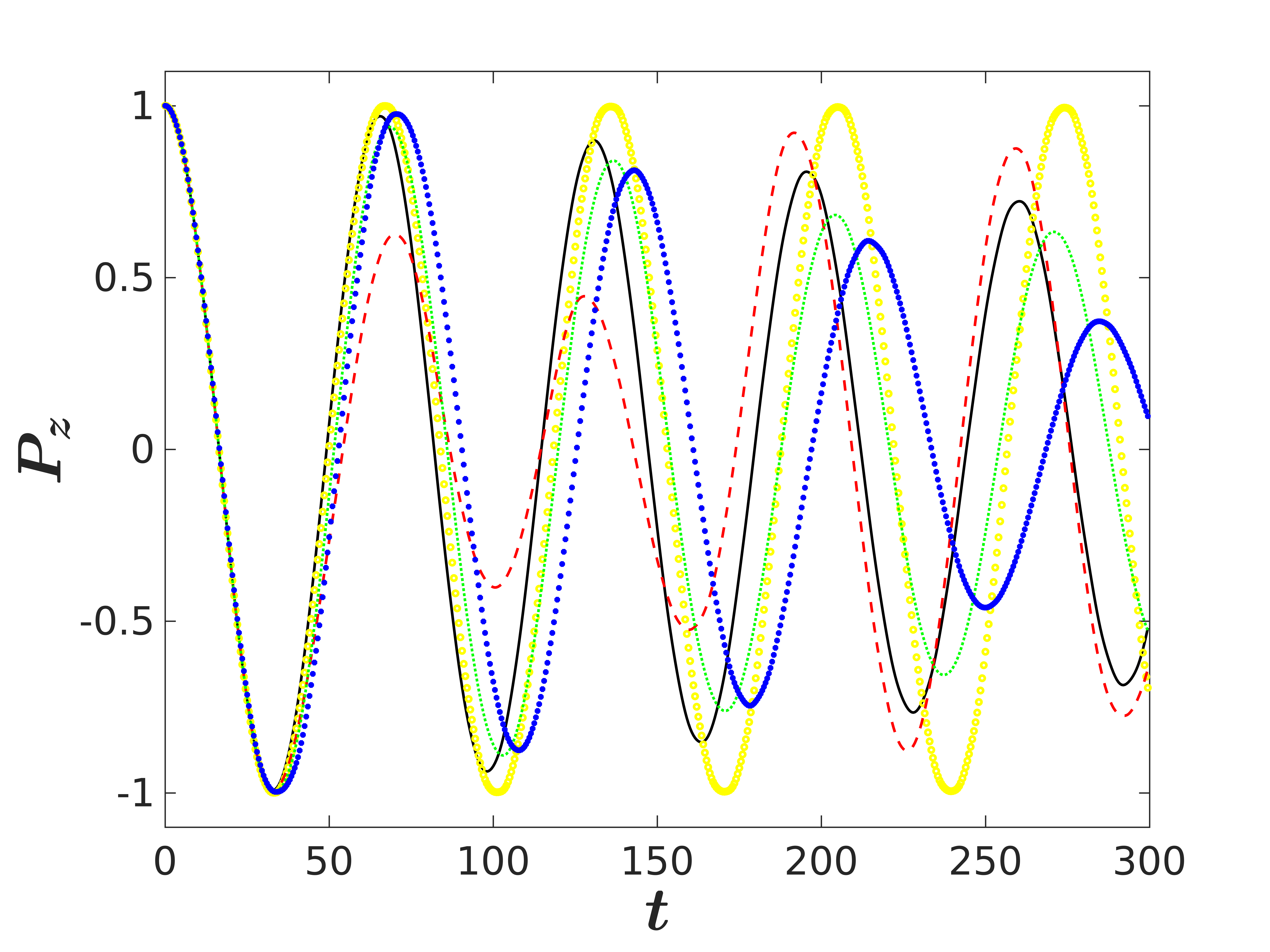}
\caption{Comparison of the time evolution of the population difference for temperature $T=1$ and system parameters $V=-0.05$ and $\lambda=0.2$:
Full quantum (full black line), stochastic Davydov (blue stars), thermally averaged Hamiltonian (yellow stars), $P$-function sampling
(green dotted line), Boltzmann averaged Davydov (red dashed line).  
\label{fig:t_1}}
\end{figure}

The constants are taken to be $\subs{k}{B}=1$, $\hbar=1$, $\omega=1$, $\varepsilon=0$.
For all methods that employ a Davydov Ansatz, the equations of motion are made explicit by dividing the equation for $\dot f$ by the respective prefactor and inserting the result into the equation for 
$\dot A$. In order to solve the resulting explicit system of ordinary differential equations, the \matlab \cite{Matlab:2015} routine \ode is applied with tolerances 
$\mbox{\texttt{relTol}} = \mbox{\texttt{absTol}} = 10^{-8}$. 

For the parameter values $V=-0.05$ and $\lambda=0.2$, the stochastic wavefunction approach of \ref{ssec:eofm} yields converged results for $N=100$ realizations. The expectation values of the right hand side 
(see \eqref{eq:stoch_expval}) are truncated for $M=7$ terms in $\ket{\Phi}$. For the $P$-function sampling of \ref{sec:PFun}, $N_s=100$ realizations for the Monte Carlo importance sampling yield converged 
results. For the Boltzmann sampling of \ref{sec:Bol}, $N_T=7$ yields converged results. For the full quantum solution outlined above, $N_T=n_{\rm max}=7$ eigenstates are propagated and the time evolution of
each of these is expanded in $M=j_{\rm max}=14$ eigenstates to achieve converged results.     


In the parameter domain where the thermally averaged Ansatz $\ket{\PsiTA}$ fails, the stochastic Davydov approach 
$\ket{\PsiT}$ still yields good agreement with the full quantum mechanical calculation for moderate temperature.
Figure \ref{fig:t_1} shows, that especially the envelope of the decay of the population density (caused by the coupling) is displayed much 
better by all the stochastic approaches than by the thermally averaged one which yields nearly constant amplitude for this case. 

\begin{figure}
\includegraphics[scale=0.3,trim= 0cm 0cm 0cm 0cm,clip=True]{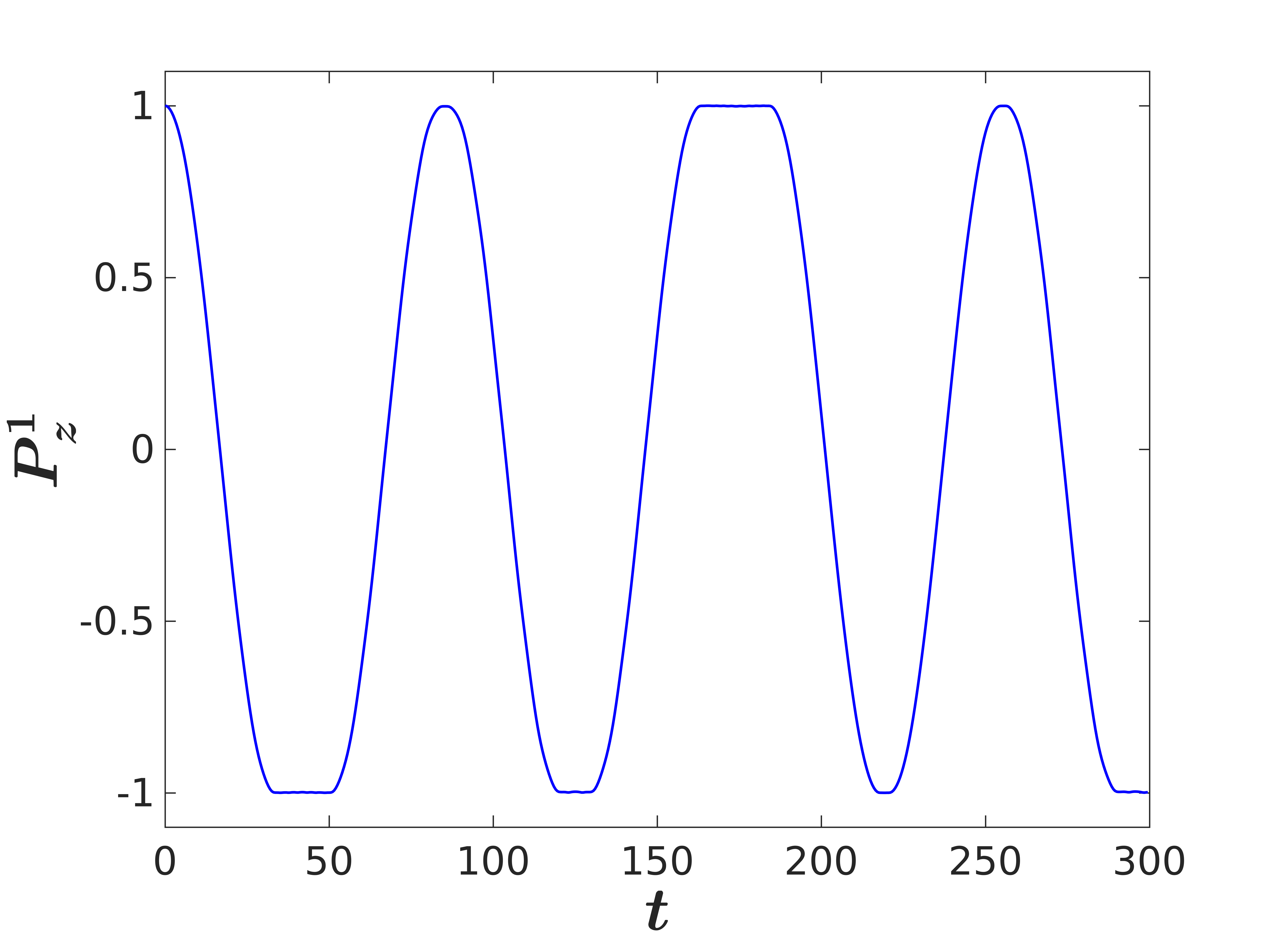}
\caption{Expectation value of $\hat{\sigma}_z$ for one single realization $n=1$, and for $V=-0.05$ and $\lambda=0.2$.  
\label{fig:t_4}}
\end{figure}

We observe bad agreement of the Davydov Boltzmann average results with the exact quantum solution. 
Taking a deeper look into the dynamics of the full quantum method shows that the temperature-dependent decay of $P_z$ results from different oscillation 
periods of the expectation values of single realizations
\be
\infsumme{j=0}\left(|c_{nj}^{(+)}(t)|^2-|c_{nj}^{(-)}(t)|^2\right), 
\ee  
although also these single expectation values decay (but at larger timescales). Also the single expectation values $P_z^n=\abs{A_n}^2-\abs{B_n}^2$ of 
the Boltzmann average do not depend on temperature. In the small coupling and small tunneling rate regime $\lambda,|V|\ll 1$, each single 
$\ket{\PsiBn(t)}$ will, while time elapses, pass through multiple critical times where $A_n(t)=0$ (resp. $B_n(t)=0$) corresponding to $P_z^n(t)=-1$ 
(resp. $P_z^n(t)=1$); and these times cannot be identified by plotting $P_z^{\rm B}$. 
Depending on the precise choice of $V$ and $\lambda$, we found that at least some of the 
single realizations show unphysical behavior near these times, 
and the larger $n$, the more often this occurs (see Figure \ref{fig:t_4})

Despite this obviously wrong behavior, energy and norm are conserved throughout. 
We attribute the mentioned problem to the fact that the explicit equations for 
$\dot f_n$ and $\dot g_n$ are singular for $A_n=0$ resp.\ $B_n=0$ due to the following reasoning: 
Firstly, the dynamics seems to be perfectly reproduced before the first minimum of $P_z$ is approached (the same holds for $P_z^n$). Secondly, 
we observe that by approaching critical times at which, e.g., $A_n=0$, singular behavior occurs when $\abs{f_n}$ gets and stays huge over relevant time-scales. 
This also contradicts the fact that $f_n$ should be arbitrary for $A_n=0$. Analytically the equation 
for $\dot f_n$ should be neglected in the case $A_n=0$ since it reduces to $0=0$ if not made explicit. But, $f_n$ cannot be set to zero 
(or any other fixed value) for $A_n=0$ since this would in general 
not result in a differentiable variable.

\begin{figure}
\includegraphics[scale=0.3,trim= 0cm 0cm 0cm 0cm,clip=True]{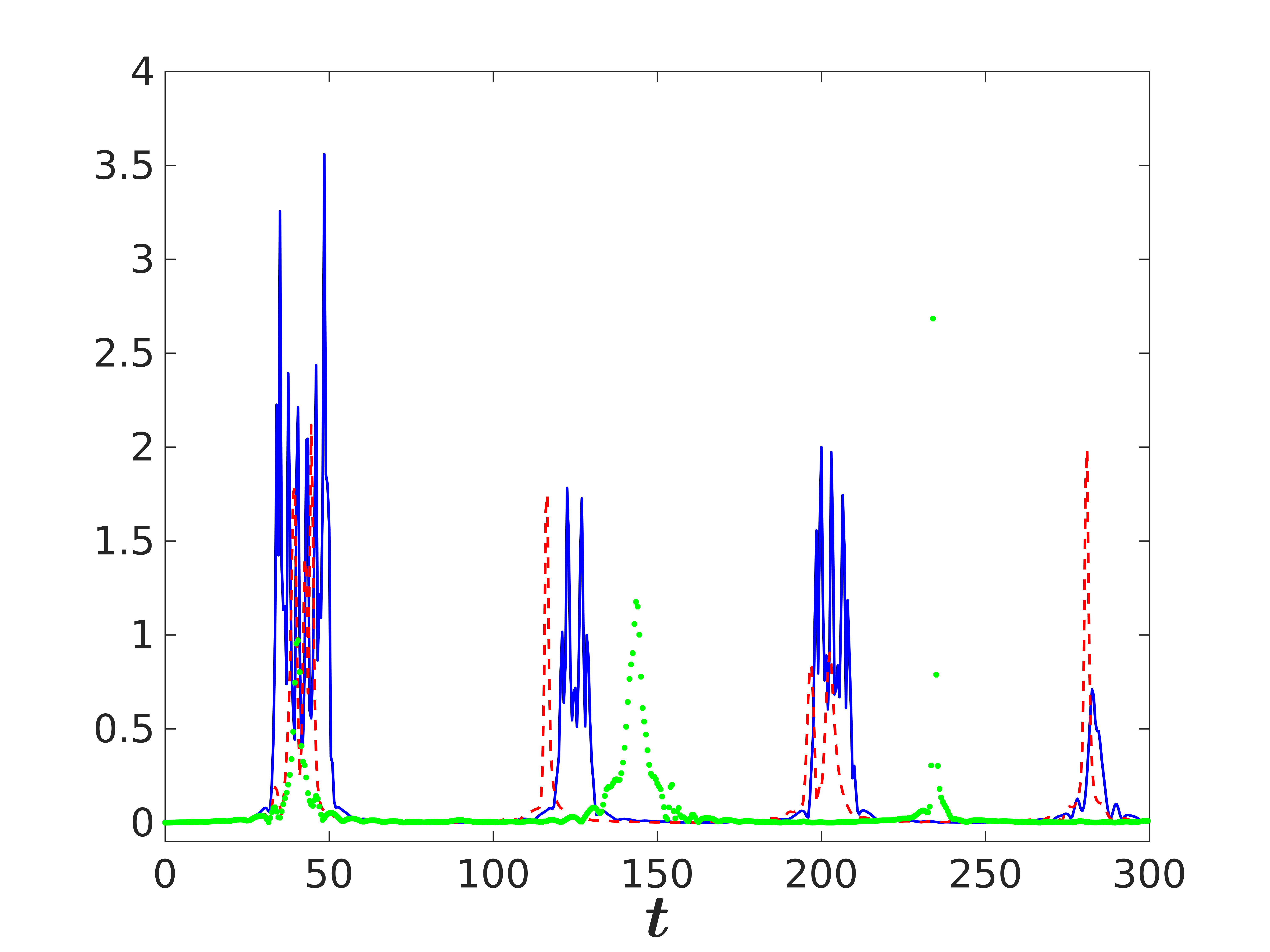}
\caption{Absolute value of the first term on the right-hand side of explicit equation for 
$\dot f_n$ for $V=-0.05$ and $\lambda=0.2$: $n=0$ (full blue line), $n=1$ (dashed red line) and $n=2$ (green stars).  
\label{fig:t_2}}
\end{figure}

To overcome the outlined difficulties, Figure \ref{fig:t_2} exemplarily shows that in the explicit equation \eqref{eq:Bol3} for $\dot f_n$,
\bea
\dot f &=& -\frac{\i V}{\hbar} {\rm e}^{-\frac 12\left(\abs{f}^2+\abs{g}^2\right)}(g-f)\left\{\frac BA {\rm e}^{f^\ast g}
\left[\laguerre_n\left(\abs{f-g}^2\right)-\laguerre_n'\left(\abs{f-g}^2\right)\right]\right. \notag \\ & & \left.\hphantom{\i\hbar\abs{A}^2\dot f 
-V{\rm e}^{-\frac 12\left(\abs{f}^2+\abs{g}^2\right)}(g-f)\left\{\right.}-\frac{B^\ast}{A^\ast} {\rm e}^{fg^\ast}\laguerre'_n\left(\abs{f-g}^2\right)\right\} \notag \\
 & & -\i\omega f-\frac{\i\lambda}{2\hbar},
\eea   
in the $|V|\ll 1$ regime the first term on the right-hand side (which is singular for $A_n=0$) is small except for $A_n\approx 0$.

Neglect of it thus seems quite natural since first we may assume that the increase of this term for $A_n\approx 0$ is a result of numerical instabilities. Furthermore this accounts for the 
assumption that the spin-system does not influence the bath, which will indeed be reasonable in the case of the bath being represented by many oscillators in the $|V|\ll 1$ regime. Even in the
extreme case of a bath of only one single oscillator,
the system \eqref{eq:Bol1}, \eqref{eq:Bol2} with simplified equations
\bea 
\dot f &=& -\i\omega f-\frac{\i\lambda}{2\hbar} \notag \\
\dot g &=& -\i\omega g+\frac{\i\lambda}{2\hbar} \label{eq:Bol_simp}
\eea
for the coherent state parameters will lead to a dramatic improvement of the numerical results as shown below. In the sequel we examine the impact on the numerics of the indicated simplification. 

Firstly, we note that initial perturbation usually imposed to overcome numerical instabilities due to $B_n(0)=0$ \cite{WFCZ17} is not needed any more. 
Secondly, the changed system still yields
\be
\frac{d}{dt}\left(\abs{A_n}^2+\abs{B_n}^2\right)=0 
\ee
resulting in norm conservation. Thirdly, we observe that for the altered as well as the unaltered system of equations, the energy of the spin system 
\be
E_s=V{\rm e}^{-\frac 12\left(\abs{f}^2+\abs{g}^2\right)}\laguerre_n\left(\abs{g-f}^2\right) 
\left[A^\ast B{\rm e}^{f^\ast g}+AB^\ast {\rm e}^{fg^\ast}\right]
\ee
is almost exactly zero. So the remaining energy
\be 
E_r=\hbar\omega\left[\abs{A}^2\left(\abs{f}^2+n\right)+\abs{B}^2\left(\abs{g}^2+n\right)\right] + \frac{\lambda}{2}\left[\abs{A}^2\left(f+f^\ast\right)-\abs{B}^2\left(g+g^\ast\right)\right]
\ee
of the bath oscillator and the coupling term is dominant (almost exactly 2 independent of time). 

It is though not constant for $g_n(0)=0$, but due to $B_n(0)=0$ we have free choice for $g_n(0)$. Since \eqref{eq:Bol_simp} can simply be solved analytically, 
we make use of 
$\abs{A_n}^2+\abs{B_n}^2=1$ to determine $g_n(0)$ such that $E_r=const.$. This results in
\be
\abs{g_n(0)-\frac{\lambda}{2\hbar\omega}}^2=\left(\frac{\lambda}{2\hbar\omega}\right)^2.
\ee
We thus set $g_n(0)=\frac{\lambda}{\hbar\omega}$, although a further phase could be chosen independently.


\begin{figure}
\includegraphics[scale=0.3,trim= 0cm 0cm 0cm 0cm,clip=True]{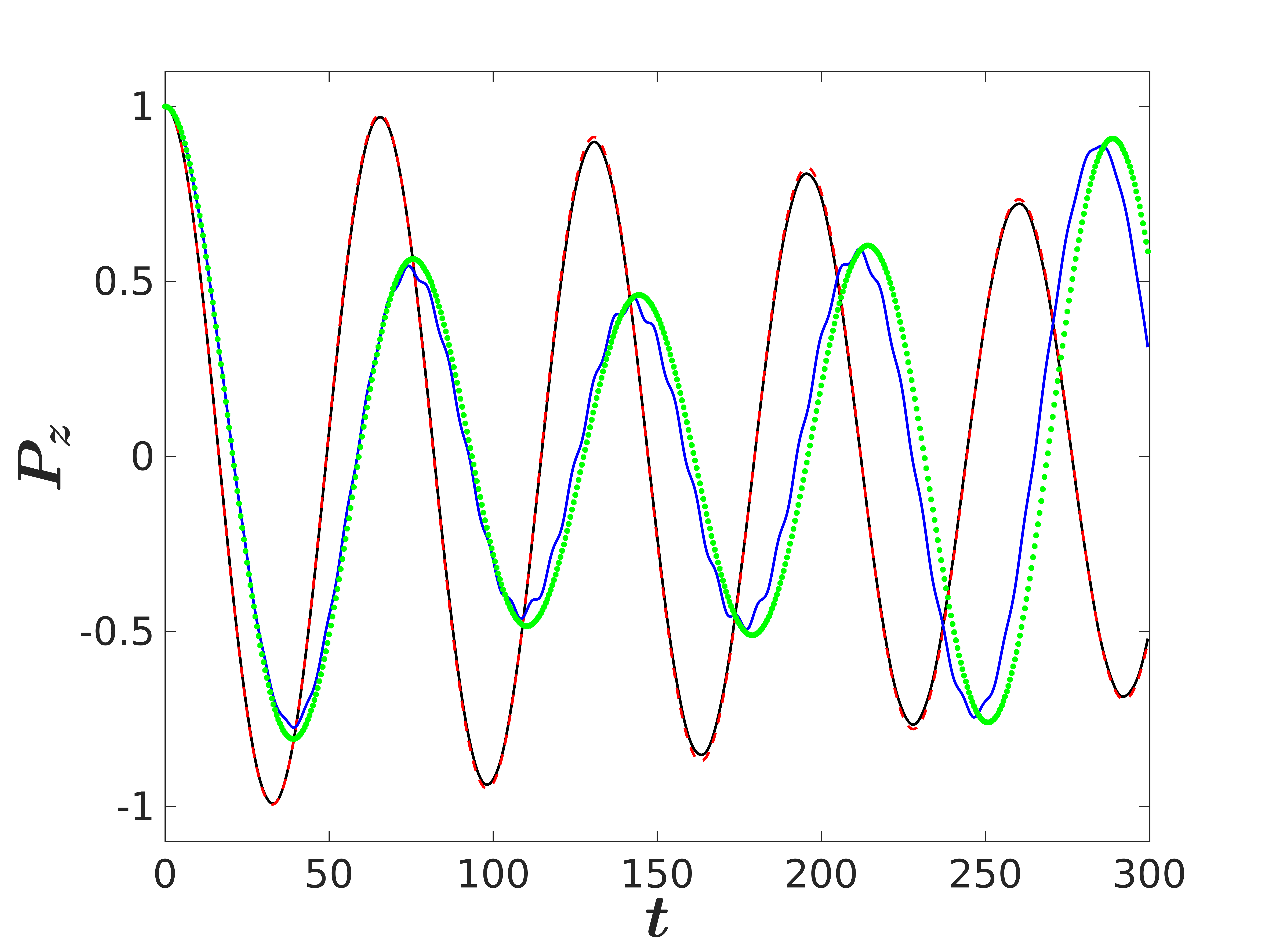}
\caption{$P_z$ for the altered equations of motion for $V=-0.05$ and $T=1$. For $\lambda=0.2$: full quantum (full black line) and 
Boltzmann averaged Davydov (red dashed line); for $\lambda=0.5$: full quantum (full blue line) and Boltzmann averaged Davydov (green stars).  
\label{fig:t_5}}
\end{figure}

Figure \ref{fig:t_5} shows perfect coincidence of the Boltzmann averaging result with simplified equations for $f$ and $g$ with the full quantum calculation for $\lambda=0.2$. 
In addition, also for $\lambda=0.5$, the interference beating is almost perfectly reproduced by the altered Boltzmann averaging method. 
In this small coupling (and small tunneling rate) regime the performed simplifications result in excellent agreement with the full quantum calculation. 
Since for the Boltzmann averaging, temperature is separate from propagation of the wave functions, the excellency of the results is independent of temperature.  

\begin{figure}
\includegraphics[scale=0.3,trim= 0cm 0cm 0cm 0cm,clip=True]{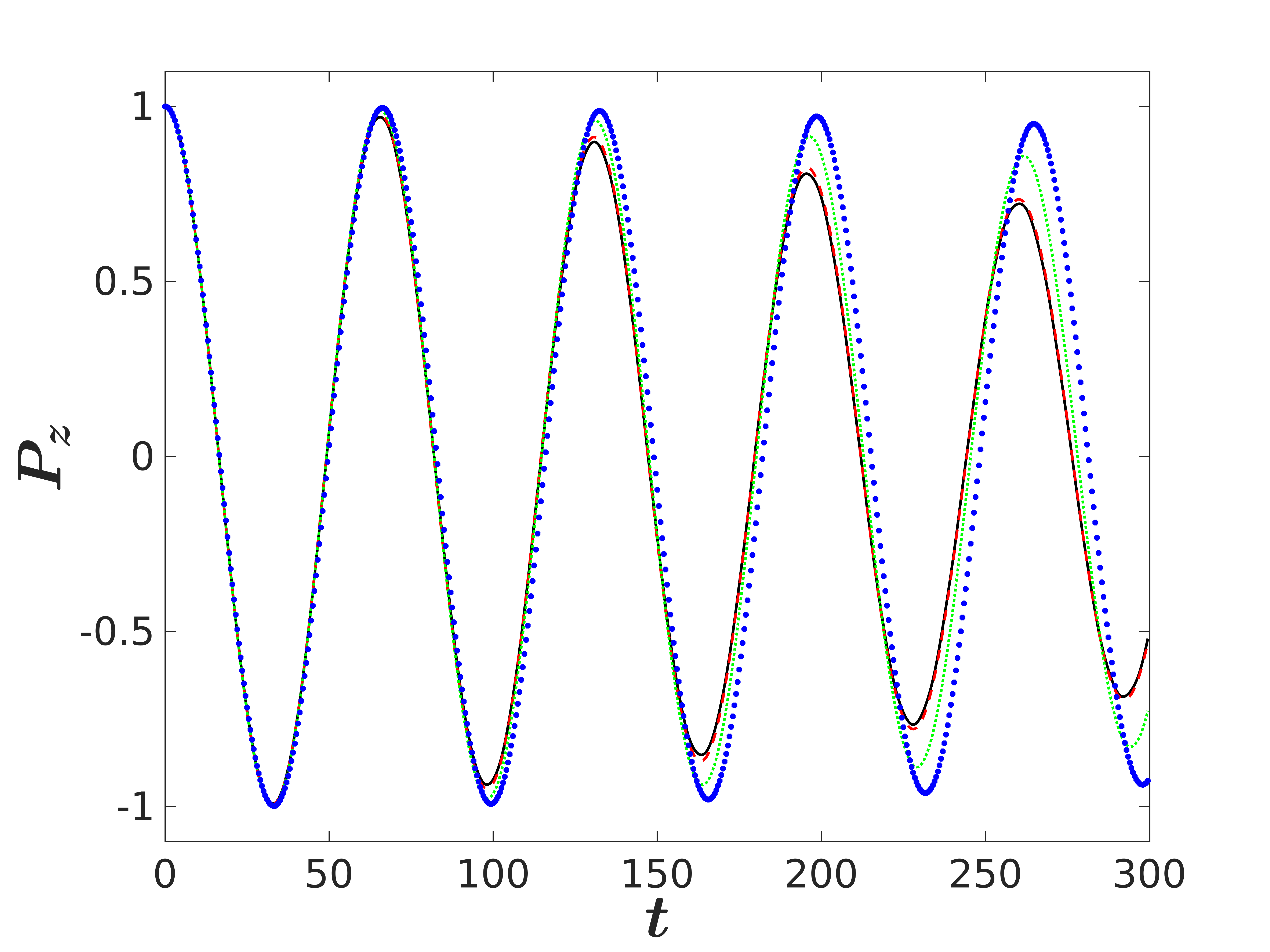}
\caption{$P_z$ for the simplified equations of motion for $V=-0.05$, $T=1$ and $\lambda=0.2$: full quantum (full black line), stochastic Davydov (blue stars), $P$-function sampling
(green dotted line), Boltzmann averaged Davydov (red dashed line).  
\label{fig:t_6}}
\end{figure}

It remains to be seen if the method outlined for the Boltzmann averaging could as well be carried out for the other methods employing a Davydov Ansatz. In the case of the $P$-function we used 
$g(0)=\alpha+\frac{\lambda}{\hbar\omega}$, and for the stochastic Davydov method $g(0)=\frac{\lambda}{\hbar\omega}$. The results of the correspondingly simplified equations of motion
can be seen in Figure \ref{fig:t_6}.
For times $t<100$ all Davydov methods show excellent agreement with the full quantum calculation. For longer times, the new stochastic method yields the worst results. This could be due to the fact 
that the equations of motion in this case still contain expressions that have to be rounded off for numerical purposes. 

\begin{figure}
\includegraphics[scale=0.3,trim= 0cm 0cm 0cm 0cm,clip=True]{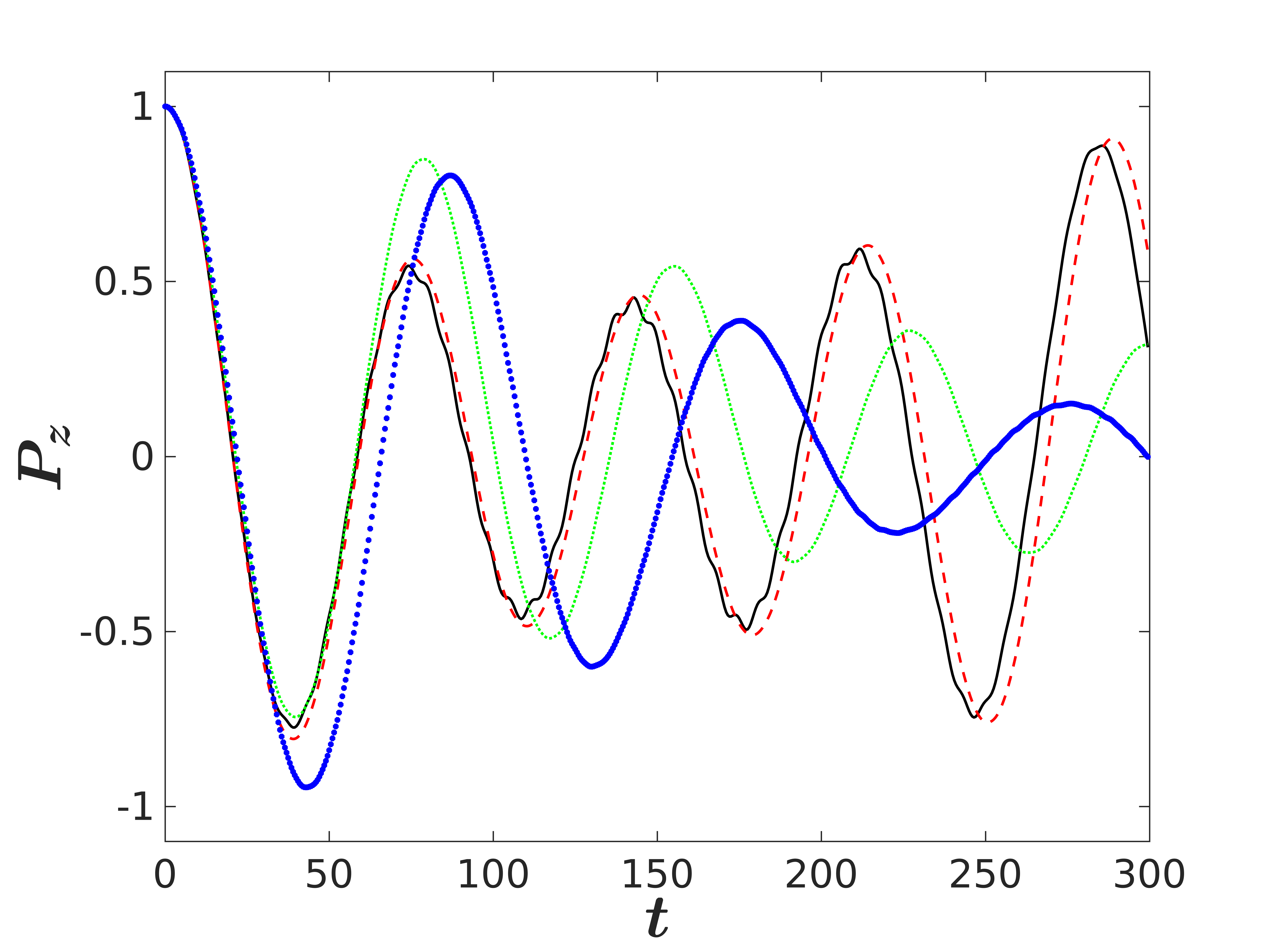}
\caption{$P_z$ for the simplified equations of motion for $V=-0.05$, $T=1$ and $\lambda=0.5$: full quantum (full black line), stochastic Davydov (blue stars), $P$-function sampling
(green dotted line), Boltzmann averaged Davydov (red dashed line).  
\label{fig:t_7}}
\end{figure}

Finally, we investigated also the case of stronger coupling $\lambda=0.5$, where a quantum beating (recurrence of the population difference) can be observed. There again, we find the best agreement  with 
the Boltzmann averaging, as shown in Figure \ref{fig:t_7}.
Both stochastic methods show a longer oscillation period compared to the full numerical quantum solution.
The new stochastic approach is again worse in this respect than the $P$-function sampling.

\section{Conclusions and Outlook}

We have compared different ways to incorporate temperature dependence into the equations of motion
following from a Davydov Ansatz for the solution of the dynamics of the quantum Rabi model. The most promising candidates are $P$-function
sampling, Boltzmann averaging and our new proposal of using a sampling scheme a la Matzkies and Manthe.

The stochastic approach for the density matrix generation a la Matzkies and Manthe that we propose is numerically
quite demanding for the case of just a single oscillator degree of freedom but may become favorable for the case of many bath degrees of freedom.
The sampling strategy using the $P$-function, with thermally sampled initial conditions instead of thermally modified equations of motion,
has already been shown to work well for the case of many bath degrees of freedom in a spin boson model with Ohmic and sub-ohmic spectral densities \cite{WFCZ17}. 
Both stochastic sampling strategies as well as the Boltzmann averaging of 
individual wavepacket propagations are superior to thermally averaged Hamiltonian propagation \cite{pccp17} as could be seen by a 
comparison to the full quantum solutions,
which are still doable for a bath consisting of just a single oscillator. Furthermore, it turned out that for the 
Davydov methods, neglecting terms that may become singular in a treatment of the explicit form of the equations leads to a dramatic improvement of the 
numerical results in the
considered parameter regime of relatively weak coupling. Furthermore, in the present single bath oscillator case, where Boltzmann averaging is
easily feasible, this approach seems superior to the two sampling strategies. A direct comparison of the sampling strategies shows a better
performance of the $P$-function method in the cases considered.

The extension of the investigations to a many mode bath with even weaker individual coupling but effectively stronger effect on the system 
will be a favorable realm for the numerical simplifications that we propose. Furthermore, 
we also stress that in general an exact solution of the time-dependent Schr\"odinger equation can be generated by using a
multi-Davydov Ansatz \cite{WFCZ17}. Also this case will be treated  in the future using our presented findings.

F.G. would like to acknowledge fruitful discussions with Profs.\ Uwe Manthe, Chang-Qin Wu and Yang Zhao. 

\bibliography{libtherm}

\begin{thebibliography}{33}
\expandafter\ifx\csname natexlab\endcsname\relax\def\natexlab#1{#1}\fi
\expandafter\ifx\csname bibnamefont\endcsname\relax
  \def\bibnamefont#1{#1}\fi
\expandafter\ifx\csname bibfnamefont\endcsname\relax
  \def\bibfnamefont#1{#1}\fi
\expandafter\ifx\csname citenamefont\endcsname\relax
  \def\citenamefont#1{#1}\fi
\expandafter\ifx\csname url\endcsname\relax
  \def\url#1{\texttt{#1}}\fi
\expandafter\ifx\csname urlprefix\endcsname\relax\def\urlprefix{URL }\fi
\providecommand{\bibinfo}[2]{#2}
\providecommand{\eprint}[2][]{\url{#2}}

\bibitem[{\citenamefont{Leggett et~al.}(1987)\citenamefont{Leggett,
  Chakravarty, Dorsey, Fisher, Garg, and Zwerger}}]{Letal87}
\bibinfo{author}{\bibfnamefont{A.~J.} \bibnamefont{Leggett}},
  \bibinfo{author}{\bibfnamefont{S.}~\bibnamefont{Chakravarty}},
  \bibinfo{author}{\bibfnamefont{A.~T.} \bibnamefont{Dorsey}},
  \bibinfo{author}{\bibfnamefont{M.~P.~A.} \bibnamefont{Fisher}},
  \bibinfo{author}{\bibfnamefont{A.}~\bibnamefont{Garg}}, \bibnamefont{and}
  \bibinfo{author}{\bibfnamefont{W.}~\bibnamefont{Zwerger}},
  \bibinfo{journal}{Rev. Mod. Phys.} \textbf{\bibinfo{volume}{59}},
  \bibinfo{pages}{1} (\bibinfo{year}{1987}).

\bibitem[{\citenamefont{Rabi}(1936)}]{Ra36}
\bibinfo{author}{\bibfnamefont{I.~I.} \bibnamefont{Rabi}},
  \bibinfo{journal}{Phys. Rev.} \textbf{\bibinfo{volume}{{\bf 49}}},
  \bibinfo{pages}{324} (\bibinfo{year}{1936}).

\bibitem[{\citenamefont{Rabi}(1937)}]{Ra37}
\bibinfo{author}{\bibfnamefont{I.~I.} \bibnamefont{Rabi}},
  \bibinfo{journal}{Phys. Rev.} \textbf{\bibinfo{volume}{{\bf 51}}},
  \bibinfo{pages}{652} (\bibinfo{year}{1937}).

\bibitem[{\citenamefont{Braak}(2011)}]{Bra11}
\bibinfo{author}{\bibfnamefont{D.}~\bibnamefont{Braak}},
  \bibinfo{journal}{Phys. Rev. Lett.} \textbf{\bibinfo{volume}{107}},
  \bibinfo{pages}{100401} (\bibinfo{year}{2011}).

\bibitem[{\citenamefont{Weiss}(2012)}]{Wei12}
\bibinfo{author}{\bibfnamefont{U.}~\bibnamefont{Weiss}},
  \emph{\bibinfo{title}{{\it{Quantum Dissipative Systems}}}}
  (\bibinfo{publisher}{World Scientific}, \bibinfo{address}{Singapore},
  \bibinfo{year}{2012}), \bibinfo{edition}{4th} ed.

\bibitem[{\citenamefont{Winter et~al.}(2009)\citenamefont{Winter, Rieger,
  Vojta, and Bulla}}]{WRVB09}
\bibinfo{author}{\bibfnamefont{A.}~\bibnamefont{Winter}},
  \bibinfo{author}{\bibfnamefont{H.}~\bibnamefont{Rieger}},
  \bibinfo{author}{\bibfnamefont{M.}~\bibnamefont{Vojta}}, \bibnamefont{and}
  \bibinfo{author}{\bibfnamefont{R.}~\bibnamefont{Bulla}},
  \bibinfo{journal}{Phys. Rev. Lett.} \textbf{\bibinfo{volume}{\bf 102}},
  \bibinfo{pages}{030601} (\bibinfo{year}{2009}).

\bibitem[{\citenamefont{Nalbach and Thorwart}(2010)}]{NT10}
\bibinfo{author}{\bibfnamefont{P.}~\bibnamefont{Nalbach}} \bibnamefont{and}
  \bibinfo{author}{\bibfnamefont{M.}~\bibnamefont{Thorwart}},
  \bibinfo{journal}{Phys. Rev. B} \textbf{\bibinfo{volume}{\bf 81}},
  \bibinfo{pages}{054308} (\bibinfo{year}{2010}).

\bibitem[{\citenamefont{Makri}(1995)}]{Mak95}
\bibinfo{author}{\bibfnamefont{N.}~\bibnamefont{Makri}}, \bibinfo{journal}{J.
  Math. Phys.} \textbf{\bibinfo{volume}{35}}, \bibinfo{pages}{2430}
  (\bibinfo{year}{1995}).

\bibitem[{\citenamefont{Stockburger}(2004)}]{St04}
\bibinfo{author}{\bibfnamefont{J.~T.} \bibnamefont{Stockburger}},
  \bibinfo{journal}{Chem. Phys.} \textbf{\bibinfo{volume}{296}},
  \bibinfo{pages}{159} (\bibinfo{year}{2004}).

\bibitem[{\citenamefont{Wang et~al.}(2001)\citenamefont{Wang, Thoss, and
  Miller}}]{WTM01}
\bibinfo{author}{\bibfnamefont{H.}~\bibnamefont{Wang}},
  \bibinfo{author}{\bibfnamefont{M.}~\bibnamefont{Thoss}}, \bibnamefont{and}
  \bibinfo{author}{\bibfnamefont{W.~H.} \bibnamefont{Miller}},
  \bibinfo{journal}{J. Chem. Phys.} \textbf{\bibinfo{volume}{115}},
  \bibinfo{pages}{2979} (\bibinfo{year}{2001}).

\bibitem[{\citenamefont{Thoss et~al.}(2001)\citenamefont{Thoss, Wang, and
  Miller}}]{TWM01}
\bibinfo{author}{\bibfnamefont{M.}~\bibnamefont{Thoss}},
  \bibinfo{author}{\bibfnamefont{H.}~\bibnamefont{Wang}}, \bibnamefont{and}
  \bibinfo{author}{\bibfnamefont{W.~H.} \bibnamefont{Miller}},
  \bibinfo{journal}{J. Chem. Phys.} \textbf{\bibinfo{volume}{115}},
  \bibinfo{pages}{2991} (\bibinfo{year}{2001}).

\bibitem[{\citenamefont{Bulla et~al.}(2005)\citenamefont{Bulla, Lee, Tong, and
  Vojta}}]{BLTV05}
\bibinfo{author}{\bibfnamefont{R.}~\bibnamefont{Bulla}},
  \bibinfo{author}{\bibfnamefont{H.-J.} \bibnamefont{Lee}},
  \bibinfo{author}{\bibfnamefont{N.~H.} \bibnamefont{Tong}}, \bibnamefont{and}
  \bibinfo{author}{\bibfnamefont{M.}~\bibnamefont{Vojta}},
  \bibinfo{journal}{Phys. Rev. B} \textbf{\bibinfo{volume}{\bf 71}},
  \bibinfo{pages}{045122} (\bibinfo{year}{2005}).

\bibitem[{\citenamefont{Grossmann}(2006)}]{jcp06}
\bibinfo{author}{\bibfnamefont{F.}~\bibnamefont{Grossmann}},
  \bibinfo{journal}{J. Chem. Phys.} \textbf{\bibinfo{volume}{125}},
  \bibinfo{pages}{014111} (\bibinfo{year}{2006}).

\bibitem[{\citenamefont{Davydov}(1980)}]{Da80russ}
\bibinfo{author}{\bibfnamefont{A.~S.} \bibnamefont{Davydov}},
  \bibinfo{journal}{Zh. Eksp. Teor. Fiz.} \textbf{\bibinfo{volume}{78}},
  \bibinfo{pages}{789} (\bibinfo{year}{1980}).

\bibitem[{\citenamefont{Kast and Ankerhold}(2013)}]{KA13}
\bibinfo{author}{\bibfnamefont{D.}~\bibnamefont{Kast}} \bibnamefont{and}
  \bibinfo{author}{\bibfnamefont{J.}~\bibnamefont{Ankerhold}},
  \bibinfo{journal}{Phys. Rev. Lett.} \textbf{\bibinfo{volume}{110}},
  \bibinfo{pages}{010402} (\bibinfo{year}{2013}).

\bibitem[{\citenamefont{Yao et~al.}(2013)\citenamefont{Yao, Duan, L\"u, Wu, and
  Zhao}}]{YDLWZ13}
\bibinfo{author}{\bibfnamefont{Y.}~\bibnamefont{Yao}},
  \bibinfo{author}{\bibfnamefont{L.}~\bibnamefont{Duan}},
  \bibinfo{author}{\bibfnamefont{Z.}~\bibnamefont{L\"u}},
  \bibinfo{author}{\bibfnamefont{C.}~\bibnamefont{Wu}}, \bibnamefont{and}
  \bibinfo{author}{\bibfnamefont{Y.}~\bibnamefont{Zhao}},
  \bibinfo{journal}{Phys. Rev. E} \textbf{\bibinfo{volume}{88}},
  \bibinfo{pages}{023303} (\bibinfo{year}{2013}).

\bibitem[{\citenamefont{Grossmann et~al.}(2016)\citenamefont{Grossmann,
  Werther, Chen, and Zhao}}]{cp16}
\bibinfo{author}{\bibfnamefont{F.}~\bibnamefont{Grossmann}},
  \bibinfo{author}{\bibfnamefont{M.}~\bibnamefont{Werther}},
  \bibinfo{author}{\bibfnamefont{L.}~\bibnamefont{Chen}}, \bibnamefont{and}
  \bibinfo{author}{\bibfnamefont{Y.}~\bibnamefont{Zhao}},
  \bibinfo{journal}{Chemical Physics} \textbf{\bibinfo{volume}{481}},
  \bibinfo{pages}{99} (\bibinfo{year}{2016}).

\bibitem[{\citenamefont{Cruzeiro et~al.}(1988)\citenamefont{Cruzeiro, Halding,
  Christiansen, Skovgaard, and Scott}}]{Cetal88}
\bibinfo{author}{\bibfnamefont{L.}~\bibnamefont{Cruzeiro}},
  \bibinfo{author}{\bibfnamefont{J.}~\bibnamefont{Halding}},
  \bibinfo{author}{\bibfnamefont{P.~L.} \bibnamefont{Christiansen}},
  \bibinfo{author}{\bibfnamefont{O.}~\bibnamefont{Skovgaard}},
  \bibnamefont{and} \bibinfo{author}{\bibfnamefont{A.~C.} \bibnamefont{Scott}},
  \bibinfo{journal}{Phys. Rev. A} \textbf{\bibinfo{volume}{37}},
  \bibinfo{pages}{880} (\bibinfo{year}{1988}).

\bibitem[{\citenamefont{Wang and Thoss}(2006)}]{WT06}
\bibinfo{author}{\bibfnamefont{H.}~\bibnamefont{Wang}} \bibnamefont{and}
  \bibinfo{author}{\bibfnamefont{M.}~\bibnamefont{Thoss}}, \bibinfo{journal}{J.
  Chem. Phys.} \textbf{\bibinfo{volume}{124}}, \bibinfo{pages}{034114}
  (\bibinfo{year}{2006}).

\bibitem[{\citenamefont{Chorošajev et~al.}(2016)\citenamefont{Chorošajev,
  Gelzinis, Valkunas, and Abramavicius}}]{CGVA16}
\bibinfo{author}{\bibfnamefont{V.}~\bibnamefont{Chorošajev}},
  \bibinfo{author}{\bibfnamefont{A.}~\bibnamefont{Gelzinis}},
  \bibinfo{author}{\bibfnamefont{L.}~\bibnamefont{Valkunas}}, \bibnamefont{and}
  \bibinfo{author}{\bibfnamefont{D.}~\bibnamefont{Abramavicius}},
  \bibinfo{journal}{Chemical Physics} \textbf{\bibinfo{volume}{481}},
  \bibinfo{pages}{108 } (\bibinfo{year}{2016}).

\bibitem[{\citenamefont{Wang et~al.}(2017)\citenamefont{Wang, Fujihashi, Chen,
  and Zhao}}]{WFCZ17}
\bibinfo{author}{\bibfnamefont{L.}~\bibnamefont{Wang}},
  \bibinfo{author}{\bibfnamefont{Y.}~\bibnamefont{Fujihashi}},
  \bibinfo{author}{\bibfnamefont{L.}~\bibnamefont{Chen}}, \bibnamefont{and}
  \bibinfo{author}{\bibfnamefont{Y.}~\bibnamefont{Zhao}}, \bibinfo{journal}{J.
  Chem. Phys.} \textbf{\bibinfo{volume}{146}}, \bibinfo{pages}{124127}
  (\bibinfo{year}{2017}).

\bibitem[{\citenamefont{Diosi et~al.}(1998)\citenamefont{Diosi, Gisin, and
  Strunz}}]{DGS98}
\bibinfo{author}{\bibfnamefont{L.}~\bibnamefont{Diosi}},
  \bibinfo{author}{\bibfnamefont{N.}~\bibnamefont{Gisin}}, \bibnamefont{and}
  \bibinfo{author}{\bibfnamefont{W.~T.} \bibnamefont{Strunz}},
  \bibinfo{journal}{Phys. Rev. A} \textbf{\bibinfo{volume}{58}},
  \bibinfo{pages}{1699} (\bibinfo{year}{1998}).

\bibitem[{\citenamefont{Borrelli and Gelin}(2016)}]{BG16}
\bibinfo{author}{\bibfnamefont{R.}~\bibnamefont{Borrelli}} \bibnamefont{and}
  \bibinfo{author}{\bibfnamefont{M.~F.} \bibnamefont{Gelin}},
  \bibinfo{journal}{J. Chem. Phys.} \textbf{\bibinfo{volume}{145}},
  \bibinfo{pages}{224101} (\bibinfo{year}{2016}).

\bibitem[{\citenamefont{Matzkies and Manthe}(1999)}]{MM99}
\bibinfo{author}{\bibfnamefont{F.}~\bibnamefont{Matzkies}} \bibnamefont{and}
  \bibinfo{author}{\bibfnamefont{U.}~\bibnamefont{Manthe}},
  \bibinfo{journal}{J. Chem. Phys.} \textbf{\bibinfo{volume}{110}},
  \bibinfo{pages}{88} (\bibinfo{year}{1999}).

\bibitem[{\citenamefont{Dirac}(1930)}]{Di30}
\bibinfo{author}{\bibfnamefont{P.~A.~M.} \bibnamefont{Dirac}},
  \bibinfo{journal}{Proc. Cambridge Philos. Soc.}
  \textbf{\bibinfo{volume}{26}}, \bibinfo{pages}{376} (\bibinfo{year}{1930}).

\bibitem[{\citenamefont{Frenkel}(1934)}]{Fren34}
\bibinfo{author}{\bibfnamefont{J.}~\bibnamefont{Frenkel}},
  \emph{\bibinfo{title}{Wave Mechanics}} (\bibinfo{publisher}{Oxford University
  Press}, \bibinfo{address}{Oxford}, \bibinfo{year}{1934}).

\bibitem[{\citenamefont{Kramer and Saraceno}(1981)}]{KS81}
\bibinfo{author}{\bibfnamefont{P.}~\bibnamefont{Kramer}} \bibnamefont{and}
  \bibinfo{author}{\bibfnamefont{M.}~\bibnamefont{Saraceno}},
  \emph{\bibinfo{title}{Geometry of the time-dependent variational principle in
  quantum mechanics}} (\bibinfo{publisher}{Springer Verlag},
  \bibinfo{address}{Berlin}, \bibinfo{year}{1981}).

\bibitem[{\citenamefont{Wu et~al.}(2013)\citenamefont{Wu, Duan, Li, and
  Zhao}}]{WDLZ13}
\bibinfo{author}{\bibfnamefont{N.}~\bibnamefont{Wu}},
  \bibinfo{author}{\bibfnamefont{L.}~\bibnamefont{Duan}},
  \bibinfo{author}{\bibfnamefont{X.}~\bibnamefont{Li}}, \bibnamefont{and}
  \bibinfo{author}{\bibfnamefont{Y.}~\bibnamefont{Zhao}}, \bibinfo{journal}{J.
  Chem. Phys.} \textbf{\bibinfo{volume}{138}}, \bibinfo{pages}{084111}
  (\bibinfo{year}{2013}).

\bibitem[{\citenamefont{F\"orner}(1992)}]{For92}
\bibinfo{author}{\bibfnamefont{W.}~\bibnamefont{F\"orner}},
  \bibinfo{journal}{J. Phys.: Condens. Matter} \textbf{\bibinfo{volume}{4}},
  \bibinfo{pages}{1915} (\bibinfo{year}{1992}).

\bibitem[{\citenamefont{Huang et~al.}(2017)\citenamefont{Huang, Wang, Wu, Chen,
  Grossmann, and Zhao}}]{pccp17}
\bibinfo{author}{\bibfnamefont{Z.}~\bibnamefont{Huang}},
  \bibinfo{author}{\bibfnamefont{L.}~\bibnamefont{Wang}},
  \bibinfo{author}{\bibfnamefont{C.}~\bibnamefont{Wu}},
  \bibinfo{author}{\bibfnamefont{L.}~\bibnamefont{Chen}},
  \bibinfo{author}{\bibfnamefont{F.}~\bibnamefont{Grossmann}},
  \bibnamefont{and} \bibinfo{author}{\bibfnamefont{Y.}~\bibnamefont{Zhao}},
  \bibinfo{journal}{Physical Chemistry Chemical Physics}
  \textbf{\bibinfo{volume}{19}}, \bibinfo{pages}{1655} (\bibinfo{year}{2017}).

\bibitem[{\citenamefont{Jeffrey and Smith}(1997)}]{JS97}
\bibinfo{author}{\bibfnamefont{S.~J.} \bibnamefont{Jeffrey}} \bibnamefont{and}
  \bibinfo{author}{\bibfnamefont{S.~C.} \bibnamefont{Smith}},
  \bibinfo{journal}{Chem. Phys. Lett.} \textbf{\bibinfo{volume}{278}},
  \bibinfo{pages}{345} (\bibinfo{year}{1997}).

\bibitem[{\citenamefont{Gardiner and Zoller}(2004)}]{GaZo}
\bibinfo{author}{\bibfnamefont{C.~W.} \bibnamefont{Gardiner}} \bibnamefont{and}
  \bibinfo{author}{\bibfnamefont{P.}~\bibnamefont{Zoller}},
  \emph{\bibinfo{title}{Quantum Noise}} (\bibinfo{publisher}{Springer-Verlag},
  \bibinfo{address}{Berlin}, \bibinfo{year}{2004}), \bibinfo{edition}{3rd} ed.

\bibitem[{Mat(2015)}]{Matlab:2015}
\emph{\bibinfo{title}{MATLAB version 8.5.0.197613 (R2015a)}},
  \bibinfo{organization}{The Mathworks, Inc.}, \bibinfo{address}{Natick,
  Massachusetts} (\bibinfo{year}{2015}).

\end{thebibliography}

\end{document}